\documentclass[sigconf]{acmart}

\AtBeginDocument{%
  \providecommand\BibTeX{{%
    \normalfont B\kern-0.5em{\scshape i\kern-0.25em b}\kern-0.8em\TeX}}}

\usepackage{booktabs}
\usepackage{float}

\usepackage[
    colorinlistoftodos,
    textsize=footnotesize,
        ]{todonotes}

\definecolor{darkgreen}{rgb}{0.0, 0.5, 0.0}

\usepackage{xspace,stfloats}

\usepackage{enumitem}
\usepackage{graphicx}
\usepackage{subfig}

\usepackage{siunitx}
\setcopyright{none}
\renewcommand\footnotetextcopyrightpermission[1]{} 

\begin{document}

\title{Analyzing Social Media Activities at Bellingcat}

\author{Dominik Bär}
\affiliation{%
  \institution{LMU Munich}
  \institution{Munich Center for Machine Learning}
  \country{Germany}
}
\email{baer@lmu.de}

\author{Fausto Calderon}
\affiliation{%
  \institution{LMU Munich}
  \country{Germany}
  }
\email{f.calderon@campus.lmu.de}

\author{Michael Lawlor}
\affiliation{%
  \institution{LMU Munich}
  \country{Germany}
  }
\email{michael.lawlor@campus.lmu.de}

\author{Sophia Licklederer}
\affiliation{%
  \institution{LMU Munich}
  \country{Germany}
  }
\email{sophia.licklederer@campus.lmu.de	}

\author{Manuel Totzauer}
\affiliation{%
  \institution{LMU Munich}
  \country{Germany}
  }
\email{m.totzauer@campus.lmu.de}

\author{Stefan Feuerriegel}
\affiliation{%
  \institution{LMU Munich}
  \institution{Munich Center for Machine Learning}
  \country{Germany}
  }
\email{feuerriegel@lmu.de}

\renewcommand{\shortauthors}{Bär, et al.}

\begin{abstract}
  Open-source journalism emerged as a new phenomenon in the media ecosystem, which uses crowdsourcing to fact-check and generate investigative reports for world events using open sources (e.g., social media). A particularly prominent example is Bellingcat. Bellingcat is known for its investigations on the illegal use of chemical weapons during the Syrian war, the Russian responsibility for downing flight MH17, the identification of the perpetrators in the attempted murder of Alexei Navalny, and war crimes in the Russo-Ukraine war. Crucial for this is social media in order to disseminate findings and crowdsource fact-checks. In this work, we characterize the social media activities at Bellingcat on Twitter. For this, we built a comprehensive dataset of all $N=$~24,682 tweets posted by Bellingcat on Twitter since its inception in July 2014. Our analysis is three-fold: (1)~We analyze how Bellingcat uses Twitter to disseminate information and collect information from its follower base. Here, we find a steady increase in both posts and replies over time, particularly during the Russo-Ukrainian war, which is in line with the growing importance of Bellingcat for the traditional media ecosystem. (2)~We identify characteristics of posts that are successful in eliciting user engagement. User engagement is particularly large for posts embedding additional media items and with a more negative sentiment. (3)~We examine how the follower base has responded to the Russian invasion of Ukraine. Here, we find that the sentiment has become more polarized and negative. We attribute this to a $\sim$13-fold increase in bots interacting with the Bellingcat account. Overall, our findings provide recommendations for how open-source journalism such as Bellingcat can successfully operate on social media.
\end{abstract}

\ccsdesc[500]{Human-centered computing~Empirical studies in collaborative and social computing}
\ccsdesc[100]{Information systems~Social networks}

\keywords{Bellingcat, Open Source Journalism,  Social Media, Twitter, Russian Invasion}

\maketitle
\pagestyle{plain}

\section{Introduction}

Open-source journalism (OSJ) presents a new phenomenon in the media ecosystem, whereby crowdsourcing and open-source information are used to create intelligence \cite{Lewis.2013}.\footnote{As the terminology is still evolving, an alternative name is, e.g., open-source intelligence (OSINT).} OSJ acts as a new journalistic practice to hold governments and other stakeholders accountable \cite{Higgins.2021}. To achieve this, OSJ collects, processes, and analyzes open-source information from social media, traditional news, and online platforms (e.g., Google Maps) with the overall goal of fact-checking and the creation of investigative reports \cite{Cooper.2021,Lewis.2013,Muller.2021}. For instance, OSJ frequently geolocates pictures based on various characteristics (i.e., houses, signs, streets, sun direction) and, thereby, verifies whether the picture is authentic or fabricated.

A prominent example of OSJ is Bellingcat (\url{http://www.bellingcat.com}). Founded by Elliot Higgins in 2014, Bellingcat is an independent, international collective of researchers, investigators, and citizen journalists from over 20 countries that uses crowdsourcing and open-source information to provide evidence on various world events \cite{Bellingcat.2022}. Among others, Bellingcat is especially known for its investigations that have confirmed the illegal use of chemical weapons during the Syrian war, the responsibility of Russia for downing flight MH17, the identification of the perpetrators in the attempted murder of Alexei Navalny, and investigations into war crimes in the Russo-Ukraine war. Beyond conflicts, Bellingcat has also investigated financial fraud, far-right movements, and human rights abuses \cite{Bellingcat.2022}.

Social media activities are an integral part of the work at Bellingcat. Specifically, interactions with their follower base occur in two main ways: (i)~information collection and (ii)~information dissemination. (i)~Information collection at Bellingcat leverages social media for crowdsourcing. In particular, Bellingcat frequently calls its followers to geolocate an image or identify the origin of specific objects. As a result, crowdsourcing efforts can provide information that are otherwise not readily available. As a prominent example, Bellingcat supported the campaign ``Stop Child Abuse'' by the European Police Office (Europol) by sharing images of objects that were used in child abuse and then asking the public to provide clues on where these objects may have originated \cite{Gonzales.2020}. (ii)~Information dissemination is important for Bellingcat to generate visibility for their own research publications, tools, and other resources, which may be useful to their community \cite{Higgins.2021}. For instance, their report on confirming a poisoning attack during the peace talks between Russia and Ukraine in April 2022 generated more than 24,000 retweets.\footnote{\url{https://twitter.com/bellingcat/status/1508463513013997580}} Since Bellingcat is not a traditional media outlet with a clear publication channel, it primarily relies on Twitter to reach a broad audience. 

So far, little is known about how OSJ operates on social media. Hence, this raises the question of how Bellingcat uses Twitter for its crowdsourcing. This is motivated by the significance of mining external knowledge for intelligence. For example, the U.S. Defense Intelligence Agency estimated that ``90 percent of worthy intelligence came from open sources [such as social media]'' \cite{Singer.2018}. To this end, this also poses the question as to how Bellingcat can elicit engagement and thus maximize the crowdsourcing capabilities of its follower base.  While there is extensive research on crowdsourcing in, e.g., crisis management \cite{Desai.2020,Gao.2011,Hunt.2019,Mulder.2016,Munro.2013,Schimak.2015}, there is a gap in our understanding of how crowdsourcing is leveraged by OSJ. We further expect that the social media activities at Bellingcat are heavily impacted by the 2022 Russian invasion of Ukraine. On the one hand, it is likely that Bellingcat may have received a broader follower base as the war is highly visible in Western countries compared to previous investigations of Bellingcat such as in Syria. This may have led to a growing audience seeking evidence. On the other hand, it may have attracted negative responses by people with pro-Russian ideology and may thus have led to a strong, negative polarization. Motivated by this, we study the following research questions (RQs):
\begin{itemize}
\item \textbf{RQ1:} \emph{How does Bellingcat use Twitter for its social media activities?}
\item \textbf{RQ2:} \emph{What are characteristics of successful social media activities at Bellingcat?}
\item \textbf{RQ3:} \emph{How did the social media activities on Bellingcat’s Twitter channel change in response to the Russian invasion of Ukraine?}
\end{itemize}

\noindent
In this paper, we aim to characterize the social media activities at Bellingcat on Twitter.\footnote{Code and data are available via \url{https://github.com/DominikBaer95/Social_media_activities_Bellingcat}} For this, we collected a comprehensive dataset of all tweets posted by Bellingcat on Twitter since its inception in July 2014 until June 2022. Overall, this amounts to $N=$~24,682 tweets. Our analysis then is three-fold: (1)~We analyze how Bellingcat uses Twitter to disseminate information and collect information from its follower base. (2)~We identify characteristics of posts that are successful in eliciting user engagement. This allows us to determine what makes their tweets go viral and, thereby, provide recommendations to successfully promote their activities. (3)~We analyze how the 2022 Russian invasion of Ukraine has impacted the social media activities at Bellingcat. Specifically, we focus on how interactions with/by the follower base have changed in response to the war. 

Our main contributions are the following: 
\begin{enumerate}[leftmargin=0.6cm] 
\item To the best of our knowledge, this is the first work characterizing the social media activities of Bellingcat as a major OSJ. 
\item We identify mechanisms with which OSJ elicits user engagement to promote information dissemination and crowdsourcing. As such, our findings provide recommendations for how OSJ can effectively operate on social media.
\item We provide evidence that the current social media activities are largely undermined by bots in response to the 2022 Russian invasion of Ukraine, thereby pointing to potential challenges in the ongoing efforts to maintain a trustworthy reputation.  
\end{enumerate}

\section{Related Work}

\subsection{Open-Source Journalism}

Open-source journalism (OSJ) emerged as a new phenomenon in the media ecosystem, which uses crowdsourcing for fact-checking and thereby generates investigative reports \cite{Cooper.2021,Lewis.2013,Muller.2021}. Such investigative reports focus, to a large extent, on conflicts around the world with the aim to promote accountability \cite{Ristovska.2022, Sienkiewicz.2014}. OSJ is grounded in the open data movement \cite{Baack.2015} and relies on the participation of various actors such as private individuals, free journalists, non-governmental organizations, and traditional media outlets \cite{Cooper.2021,Mutsvairo.2022}. To gather data, OSJ crowdsources various information from open sources. Examples are social media, traditional news, and online platforms (e.g., Google Maps, flight trackers, weather services), but information sources may also include leaked company databases. 

Existing literature has approached OSJ primarily through a qualitative lens with the aim of better conceptualizing the phenomenon. Specifically, scholars have highlighted the value of open-source practices for investigative journalism and stress democracy and accountability as a main source of motivation \cite{Muller.2021, Sienkiewicz.2014}. Consequently, previous research has emphasized the importance of OSJ for investigations in war and conflict \cite{Baack.2015}. Here, OSJ can provide conclusive evidence using open data sources and crowdsourcing \cite{Hauter.2022, Sienkiewicz.2014}. However, these works are all purely qualitative and do not provide a quantitative analysis of OSJ characteristics.

\subsection{Bellingcat} 

A prominent example of OSJ is Bellingcat (also spelled bell{?`}ngcat). Founded by Elliot Higgins in 2014, Bellingcat now has regular contributors in more than 30 countries \cite{Bellingcat.2022}. Bellingcat is particularly known for its investigations confirming (1)~the illegal use of chemical weapons during the Syrian war,\footnote{\url{https://www.bellingcat.com/news/mena/2018/06/14/chemical-munitions-used-syrian-government-2012-2018/}} (2)~the responsibility of Russia for downing flight MH17,\footnote{\url{https://www.bellingcat.com/news/uk-and-europe/2015/10/08/mh17-the-open-source-evidence/}} (3)~the identification of the perpetrators in the attempted murder of Alexei Navalny,\footnote{\url{https://www.bellingcat.com/news/uk-and-europe/2020/12/14/fsb-team-of-chemical-weapon-experts-implicated-in-alexey-navalny-novichok-poisoning/}} and (4)~investigations into Russian war crimes in Ukraine.\footnote{\url{https://www.bellingcat.com/tag/ukraine/}} Crucial for this is social media, which allows Bellingcat to disseminate findings but also to crowdsource fact-checks \cite{Bellingcat.2022, Cooper.2021,Muller.2021}, implying that information between Bellingcat and their follower base can flow in both ways. 
To this date, only few works have studied the activities of Bellingcat. There is one work documenting how Bellingcat investigated the downing of flight MH-17 in 2014 \cite{Ilyuk.2019}. This work argues that the key enabler for this was to reconstruct the timeline of the Russian military transporting a missile system from St. Petersburg to the Ukrainian border and, afterward, separatists launching a Buk missile near Snizhne, Donetsk Oblast. However, to the best of our knowledge, there is no research characterizing the social media activities of Bellingcat. 

\subsection{Journalism in the Social Media Era}

Social media serves as a major news source for large parts of modern societies \cite{Walker.2021}. As a result, many traditional media outlets are active on social media platforms such as Facebook and Twitter. In this regard, previous research has studied how media outlets use social media platforms to share news \cite{Aldous.2019,Engesser.2015,Humayun.2022,Russell.2019}. Here, results show that media outlets (e.g., newspapers) have different usage patterns as compared to individual journalists \cite{CarrascoPolaino.2022,Humayun.2022,Orellana-Rodriguez.2016}. For example, one finding is that media outlets use Twitter frequently, though not skillfully. Moreover, many media outlets hardly interact with their audience and barely use characteristic elements of social media such as hashtags \cite{Engesser.2015}. Instead, media outlets typically use social media to link to content on their websites or retweet articles by their journalists \cite{Russell.2019}. In contrast to media outlets, individual journalists frequently make use of social media not only for self-promotion but also to collect information \cite{Humayun.2022}. This is similar to OSJ such as Bellingcat that, different from traditional media outlets, leverages followers for information collection and verification, and thus heavily uses bidirectional communication on social media. 

Previous research has also studied how individuals engage with news content on social media. One stream in the literature has shown that, besides the information value of news content, user engagement with news content on social media depends on \emph{who} is posting \cite{Orellana-Rodriguez.2016}. For example, individuals prefer individual journalists more than organizations because of the personal relation \cite{CarrascoPolaino.2022}.

Another stream of literature studies the effect of \emph{what} is posted on user engagement, such as views, likes, replies, and reposts. For example, content characteristics (e.g., hashtags, mentions, URLs, and images) can be used to drive user engagement \cite{CarrascoPolaino.2022,Hwong.2017,Naveed.2011,Robertson.2022}. Specifically, posts with hashtags, mentions, and URLs have been found to be more likely to be retweeted and thus go viral \cite{Naveed.2011}. Findings also suggest that news posts are more engaging when they include images \cite{Hwong.2017}. Additionally, one study confirms that the inclusion of URLs and hashtags proliferates the virality of social media content \cite{Suh.2010}. Further, user engagement varies across topics with protests being especially engaging \cite{Aldous.2019b}. However, these findings are limited to traditional media outlets, whereas we analyze how OSJ uses social media and further identify characteristics of successful social media activities to elicit user engagement. 

\subsection{Crowdsourcing}

Crowdsourcing refers to practices whereby a job is delegated to a large group of individuals, typically through an open call \cite{Desai.2020, Howe.2008, Mulder.2016}. As such, crowdsourcing is an iterative process where multiple parties collect and evaluate information, thereby reaching high levels of accuracy, especially, compared to official sources \cite{Meier.2012, Schimak.2015}. Prior research has studied crowdsourcing in various settings, such as for crisis management \cite{FloresSaviaga.2021, Gao.2011}, prediction markets \cite{Freeman.2017}, or funding \cite{Lu.2014}, whereas we add by crowdsourcing for investigative intelligence. 

\vspace{0.1cm}
\noindent
\textbf{Research gap:} To the best of our knowledge, no work has studied the social media activities of OSJ. To close this gap, we (1)~analyze how Bellingcat, as a prominent example of OSJ, uses Twitter for its social media activities, (2)~identify characteristics of successful social media activities for OSJ, and (3)~evaluate how Bellingcat's social media activities have changed in response to the Russo-Ukraine war. Thereby, we make recommendations for how OSJ can successfully leverage social media. 

\section{Data Collection}

We study the social media activities at Bellingcat due to its prominence for OSJ. For this, we collected all tweets posted by Bellingcat on Twitter from its inception in July 2014 through June 4, 2022. Here, we focus on Twitter as it is Bellingcat's main online communication channel with more than 700,000 followers. For comparison, the follower base at both Facebook: ($\sim$81,000 followers) and Telegram ($\sim$31,000 members) are substantially smaller in size, and both platforms receive less comprehensive coverage as only a subset of Twitter content is posted there. Overall, we obtained $N=$~24,862 tweets in English from Bellingcat via Twitter's REST API.\footnote{https://developer.twitter.com/en/docs/twitter-api} This includes tweets, quoted tweets, replies, and retweets. We further collected all replies to Bellingcat's tweets. We restricted our data collection to English tweets as we are interested in studying the bidirectional information flows and this is the language used by Bellingcat. Overall, our dataset contains 64,525 replies, out of which 15,080 were posted after the Russian invasion.

In our data, we distinguish tweets and threads. A thread refers to several tweets posted at once by an author to extend the 280-character limit and thereby allows to structure longer posts. Bellingcat's communication on Twitter largely depends on threads to add additional context and information on their investigations. Hence, unless stated otherwise, all our analyses below are at the thread-level. Single tweets are also considered a thread such that no information is discarded. In total, Bellingcat has posted 7,345 threads in English of which 258 were posted in the 100 days before and after the Russian Invasion of Ukraine.

\section{Methods}
\label{sec:methods}

In the following, we provide methodological details for our (1)~content analysis and (2)~bot identification. 

\subsection{Content Analysis}

We performed a content analysis to understand (1)~\emph{what} content is shared, (2)~\emph{how} positively/negatively the content is phrased, and (3)~\emph{why} the content has been shared (i.e., the purpose). For this, we used different methods, namely topic modeling (what), sentiment analysis (how), and a qualitative assessment (why), respectively. 

\textbf{Topic modeling:} To analyze \emph{what} content is shared by Bellingcat, we applied a latent Dirichlet allocation (LDA) \cite{Blei.2003} to infer meaningful topics from Bellingcat's threads. We chose LDA as it has been used in previous research studying user engagement by topics across different social media platforms \cite{Aldous.2019b}. We also considered the use of a BERT-based clustering for topic modeling \cite{Toetzke.2022} but eventually discarded this. The reason was that our corpus contains slang that is highly informative. In particular, our corpus has many hashtags, acronyms, and other domain-specific vocabulary that would result in out-of-vocabulary tokens when using BERT and thus would have been lost.  

Before applying LDA, we removed stopwords, performed lemmatization, and tokenized all threads. We set the number of topics ($k$) to 10. We also experimented with other choices (i.e., $k=1$ to $k=20$) for which we analyzed topic coherence. Specifically, topic coherence increased rapidly for values up to $\sim8$ topics but showed signs of saturation for $\sim10$ topics. Upon manual inspection, we found that 10 topics were able to discover sufficiently diverse themes without being too granular. Afterward, we manually assigned names to each topic by inspecting the top~10 words per topic. We used the LDA implementation in scikit-learn. The maximum iteration was set to 200. This was higher than the default value and was used to ensure sufficient convergence. 

\textbf{Sentiment analysis:} To determine \emph{how} positive/negative content is, we extracted the sentiment of each thread by Bellingcat using VADER \cite{Hutto.2014}. VADER is a lexicon-based approach for sentiment analysis, which has been specifically developed for texts from microblogs such as Twitter \cite{Hutto.2014}. Furthermore, lexicon-based approaches have been frequently used in prior works to analyze social media data \cite{Bonta.2019}, e.g., to study the spread of online rumors \cite{Prollochs.2021, Prollochs.2021b} or to characterize conspiracy theorists \cite{Bar.2022}. As a robustness check, we also performed sentiment analysis using the NRC dictionary \cite{Mohammad.2021b} and LIWC 2015 \cite{Pennebaker.2015}, which led to qualitatively the same results. For each thread, VADER returns a sentiment score between $-1$ and $+1$, where large values refer to a more positive sentiment. For reasons of interpretability, we discretized the scale according to the thresholds recommended by \cite{Hutto.2014}. Accordingly, we classify threads with sentiment values $<-0.5$ as ``negative'', $>0.5$ as ``positive'', and otherwise as ``neutral''.

\textbf{Manual classification by purpose:} To identify the purpose of a thread posted by Bellingcat (\emph{why}), we manually coded threads using the methodology outlined by \cite{Ryan.2003}. By comparing the differences and similarities of all threads, we identified 6 purpose categories (see Tbl.~\ref{tbl:thread_purpose}). Subsequently, two authors independently assigned each thread to one purpose category. Here, we find a statistically significant inter-rater reliability in
terms of Cohen’s kappa $\kappa = 0.53$ which implies a moderate agreement according to standard reporting guidelines \cite{Landis.1977}. In the next step, the authors followed previous research \cite{CavazosRehg.2016}, and iteratively discussed and relabeled the threads until they converged in a single purpose category for each thread. During this process, the authors revised the definitions for the categories \textsc{Bellingcat Ops} and \textsc{Self-Promotion} to resolve disagreement and clearly distinguish promotional and operational purposes. Accordingly, we used \textsc{Self-Promotion} to refer exclusively to content for advertisement and promotion purposes, while \textsc{Bellingcat Ops} refers to content regarding Bellingcat's organization such as hiring, finances, legal issues, or other operational aspects.

\begin{table*}[h!]
\centering
\footnotesize
\begin{tabular}{lp{5cm}p{9cm}  }
\toprule
Category & Definition  & Example \\
\midrule
\textsc{Research Publication} & Twitter threads about the presentation of journalistic stories or findings from Bellingcat’s work. & \emph{Fascist Fashion: How Mainstream Businesses Enable the Sale of Far-Right Merchandise https://t.co/VBp2X8vlRH} \\
\midrule
\textsc{Crowdsourcing}  & Threads that ask for help or collaboration from the public for a certain task. & \emph{ We are looking for videos and photographs from Okhtyrka  related to any munitions or attack sites that have been documented today, and also ask for assistance geolocating any of those images. Where possible please provide the source. https://t.co/14dCjxsFO4}  \\
\midrule
\textsc{Bellingcat Ops} & Threads about an event related to the organization itself. This includes topics regarding Bellingcat's staff, finances, legal issues, or operations. & \emph{Russia's Federal Service for Supervision of Communications, Information Technology and Mass Media (Roskomnadzor) has reportedly banned Bellingcat: https://t.co/qTOL4T2OH9}   \\
\midrule

\textsc{Self-Promotion} & Threads that advertise, their content, work, or appearances. & \emph{Join @GoodLawProject, @bellingcat + @carolecadwalla on March 25th for an online discussion on the power of open-source investigations and citizen journalism to hold those in power to account. https://t.co/yhUe9gEi3a}   \\
\midrule
\textsc{Other-Promotion} & Threads that share other accounts' or websites' content. & \emph{Concerns over radiation at Chernobyl, hopefully @Maxar or @Planet can get an image that could show damage in the area. https://t.co/NlmaL8NLqh}\\
\midrule
\textsc{Tools and Training} & Threads that present tools, resources, or training with educational purposes on OSJ. & \emph{Our interactive TimeMap feature detailing incidents of civilian harm in Ukraine should now be compatible with mobile platforms. We'll continue to update our findings as the conflict progresses. You can view the TimeMap on desktop or mobile here: https://t.co/h0ZdeTjbV0 https://t.co/Q6LcBGQmhU} \\ 
\bottomrule
\end{tabular}
\caption{Coding scheme for classifying content by purpose.}
\label{tbl:thread_purpose}
\end{table*}

\subsection{Bot Identification}

We identified bots using Botometer \cite{Sayyadiharikandeh.2020}. Botometer is a supervised machine learning classifier assessing the likelihood of an account being a bot and, to do so, considers over 1,000 different features such as an account's metadata, friendship network, and activity patterns. Botometer exhibits a high accuracy (0.96 AUC \cite{Sayyadiharikandeh.2020}) and has been frequently used in previous works \cite{Geissler.2022, Shao.2018,Wojcik.2018}.

We accessed Botometer via the public API (\url{botometer.osome.iu.edu}) maintained by the Indiana University Observatory on Social Media. For each account, it returns the probability of being a bot. In line with previous research \cite{Shao.2018}, we classified accounts with Botometer scores $>0.5$ as bots.

\section{Results}

\subsection{RQ1: How does Bellingcat use Twitter for its social media activities?}

To analyze how Bellingcat uses Twitter, we focus on the following five dimensions: (1)~the temporal evolution of followers and interactions; (2)~the posting activity; (3)~the use of media items, hashtags, and mentions; (4)~topics in social media threads for information collection, information dissemination, and other activities; and (5)~sentiment embedded in threads. This allows us to shed light on Bellingcat's social media strategy.

\begin{figure}
    \centering
    \subfloat[\centering Followers]{{\includegraphics[width=.85\linewidth]{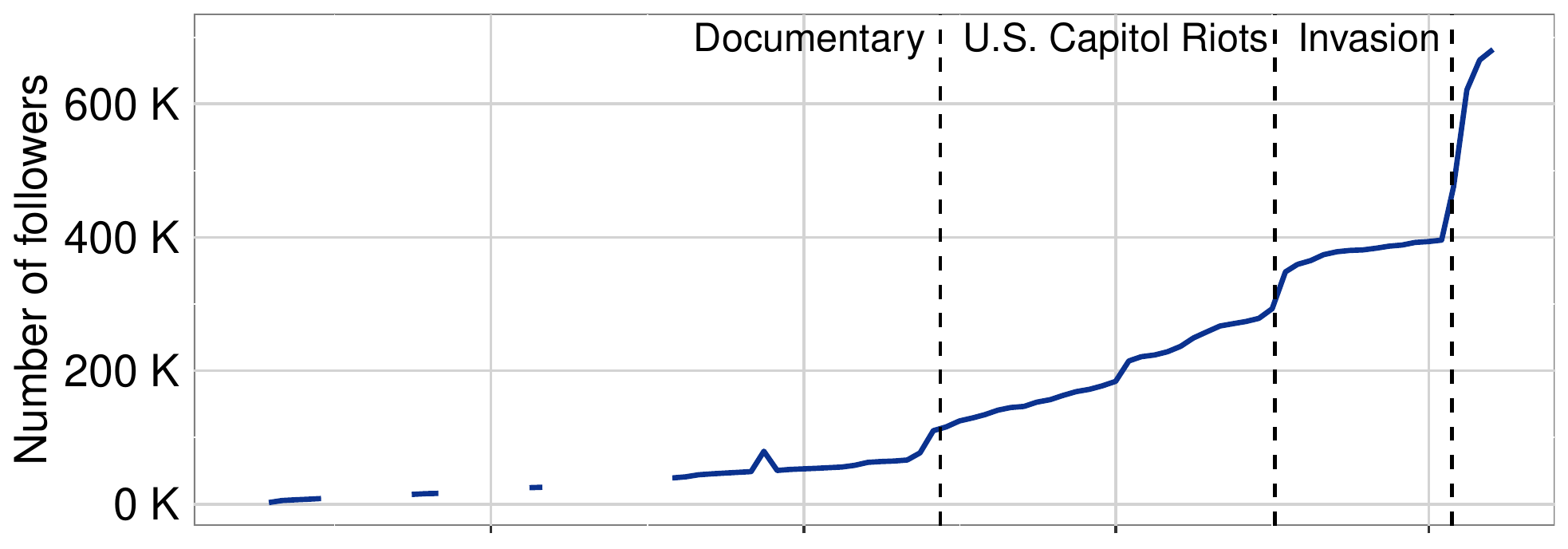}}}%
    \qquad
    \subfloat[\centering Interactions]{{\includegraphics[width=.85\linewidth]{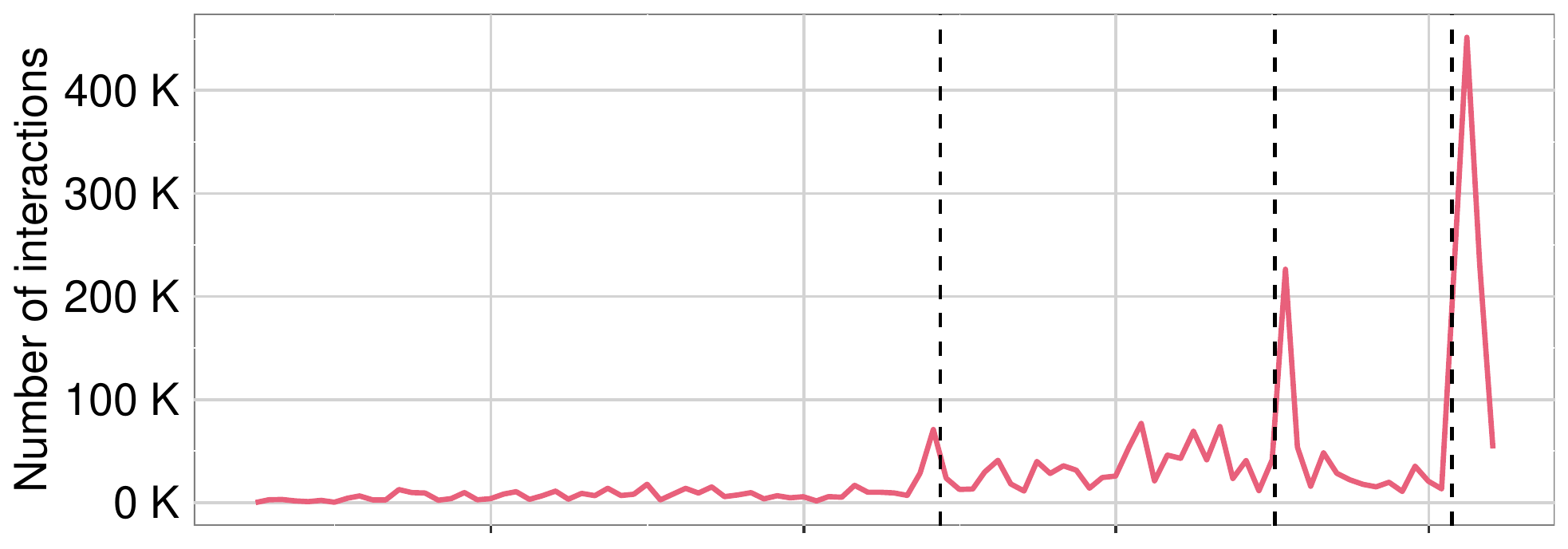}}}%
     \qquad
    \subfloat[\centering Posts]{{\includegraphics[width=.85\linewidth]{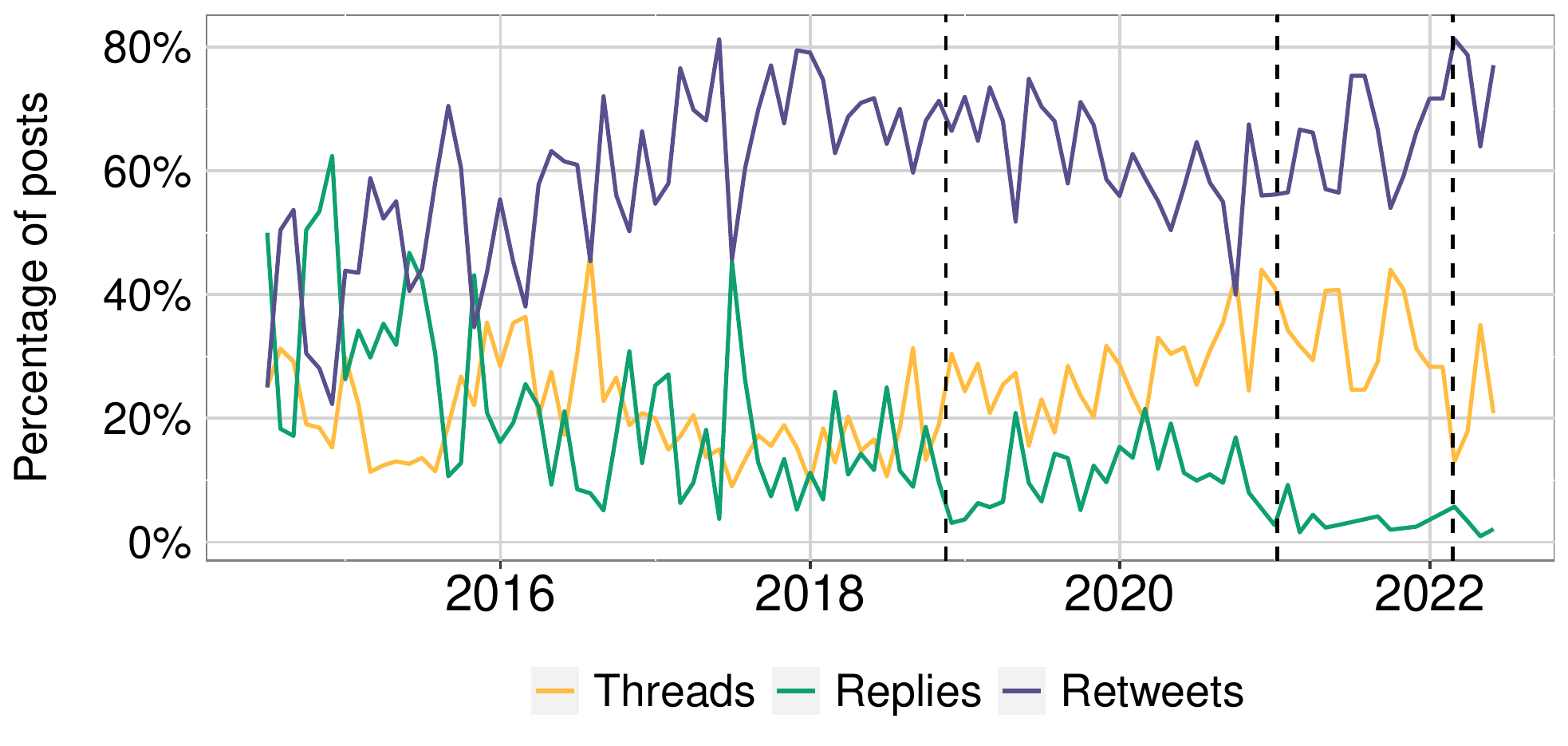}}}%
    \caption{Number of (a)~followers, (b)~interactions, and (c)~posts per month. Vertical lines denote impactful events.}%
    \label{fig:followers_interactions_posts}%
\end{figure}

\vspace{0.1cm} \noindent
\emph{How do Bellingcat's followers and interactions develop over time?} To better understand Bellingcat's growth on Twitter, we first proceed by examining the follower count over time; see Fig.~\ref{fig:followers_interactions_posts}(a). To do so, we retrieved historical data on the  monthly number of followers to Bellingcat's Twitter account via \url{WebArchive.org}. We find that the follower base is heavily impacted by three key events, namely the release of the Bellingcat documentary in November 2008, investigations around the U.S. Capitol riots in January 2021, and investigations on the Russian invasion of Ukraine in February 2022. Specifically, after these events, Bellingcat's follower count increased by $\sim$43\,\%, $\sim$19\,\%, and $\sim$20\,\%, respectively. Hence, investigations of important, global events may present an effective way to further steer growth in the follower base and achieve a broader audience.

The total number of interactions (i.e., the sum of retweets, replies, quotes, and likes from other users to Bellingcat's threads) shows a similar pattern; see Fig.~\ref{fig:followers_interactions_posts}(b). In particular, the number of interactions peaked after the release of the documentary, the U.S. Capital Riots, and the Russian invasion of Ukraine. Overall, these events led to an $\sim$149\,\%, $\sim$445\,\%, and $\sim$1536\,\% increase in interactions, respectively. Hence, this identifies offline dynamics as an important driver for interactions. However, the effect of such offline events is rarely persistent and levels out quickly. 

\vspace{0.1cm} \noindent
\emph{How does Bellingcat's Twitter activity evolve?} Bellingcat's own activity on Twitter is shown in Fig.~\ref{fig:followers_interactions_posts}(c). We find that Bellingcat's use of threads remains equally important throughout the years. For example, the share of threads (out of all posts) amounts to 26\,\%, 17\,\%, and 20\,\% in 2014, 2018, and 2022, respectively. This can be expected as threads allow Bellingcat to share investigations. We also find changes in the posting strategy, as Bellingcat's replies decrease over time: While replies account for 34\,\%  of all posts by Bellingcat in 2014, they only account for 15\,\% of all posts in 2018 and a mere  3\,\% in 2022. Hence, Bellingcat tends to engage less in conversations with its audience. In contrast, the share of retweets grew over time and accounts for 40\,\%, 68\,\%, and 76\,\% of all posts in 2014, 2018, and 2022, respectively. A reason for this is that Bellingcat frequently retweets posts from personal accounts owned by members of the organization (e.g., Elliot Higgins, Christo Grozev, Giancarlo Fiorella) or associated organizations (e.g., Forensic Architecture, Centre for Information Resilience). 

\vspace{0.1cm}\noindent
\emph{What strategy does Bellingcat employ to use media items, hashtags, and mentions?} A common characteristic of Twitter content is the frequent use of media items (i.e., URLs, images, and videos), hashtags, and mentions, which allows one to invite others to join the conversation. By analyzing these dimensions, we gain a better understanding of how threads are designed. 

\begin{figure}[H]
\centering
\includegraphics[width=0.75\linewidth]{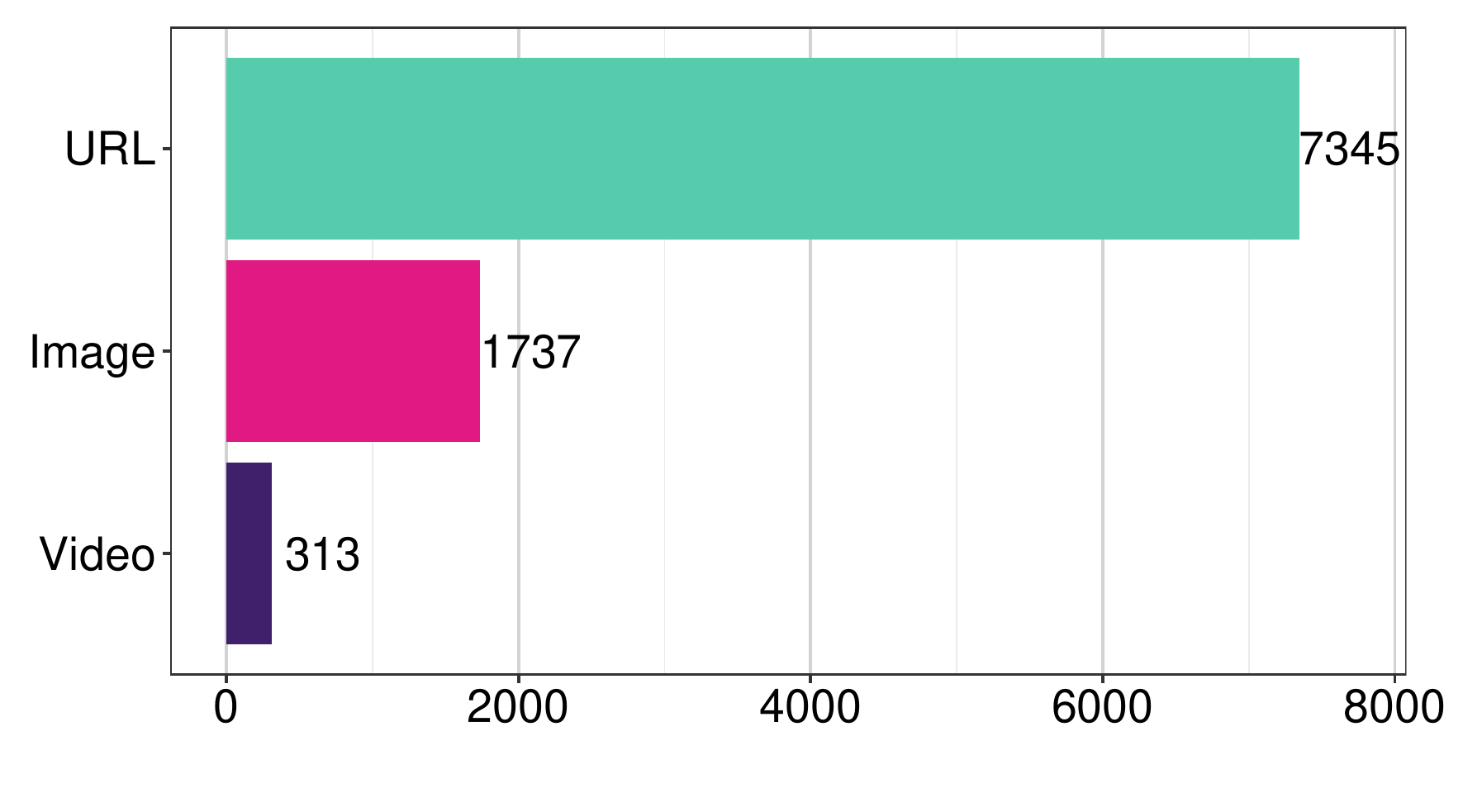}
\caption{Frequency of media items and URLs.}
\label{fig:freq_media_items}
\end{figure}

Fig.~\ref{fig:freq_media_items} shows the distribution of different media items. Overall, 93\,\% of Bellingcat's threads feature additional media items. For example, 7,345 (97\,\%) of Bellingcat's threads included URLs. The large number of URLs shows that Bellingcat frequently references related content or promotes its own investigations on its website. Furthermore, Bellingcat shared 1,737 images along with their threads. Here, images may help Bellingcat for two reasons: (1)~Images can increase the reach of content on Twitter \cite{Li.2020b} and thus may help to inform more people about its investigations. (2)~The use of images allows Bellingcat to support their investigations with visual information, thus making them more convincing for the audience \cite{Russell.2019}. In contrast, videos are comparatively rare (4\,\% of all threads). 

\begin{figure}[H]
    \centering
    \subfloat[\centering Hashtags]{{\includegraphics[width=0.49\linewidth]{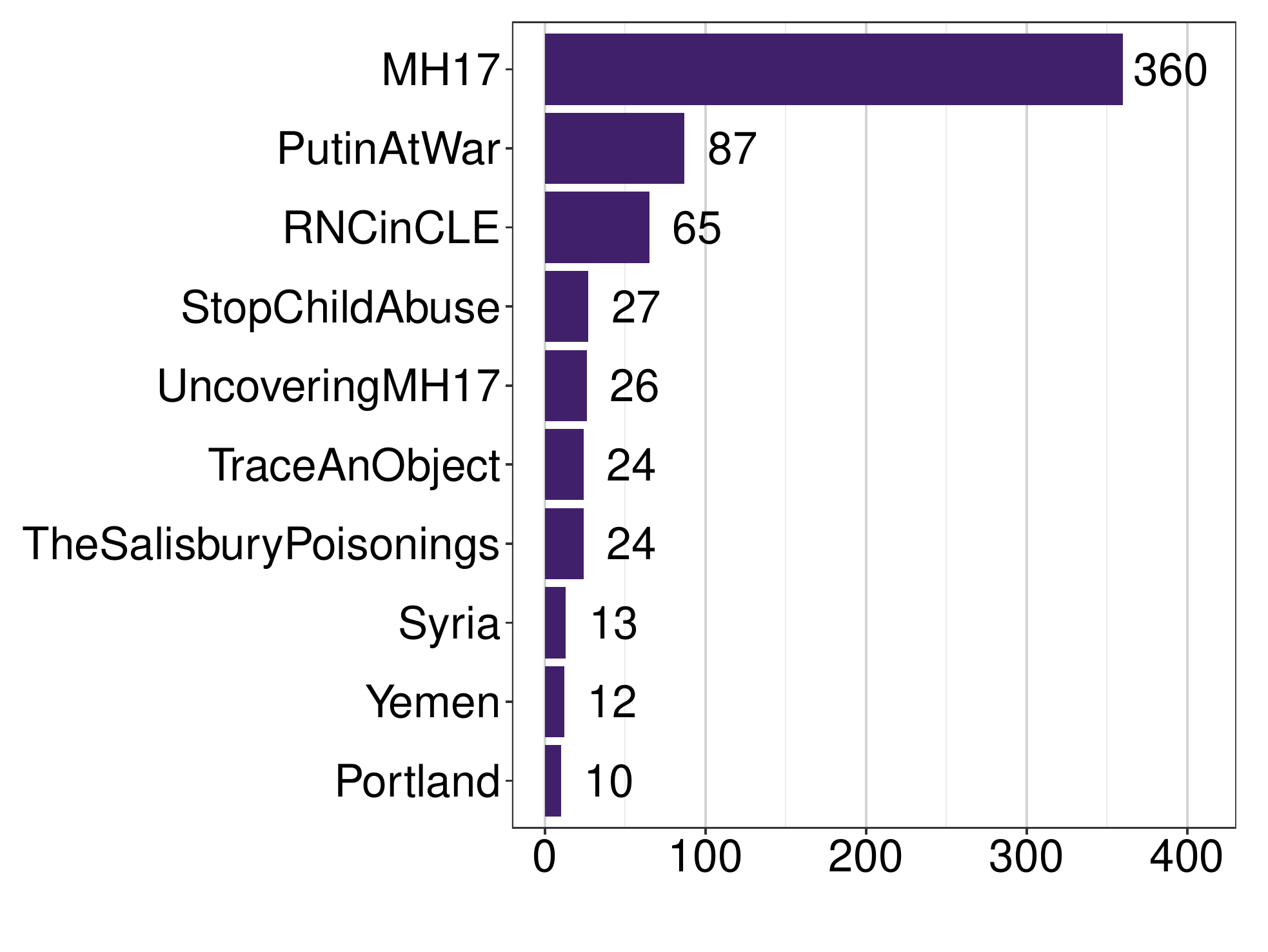}\hspace{-2em} }}%
    \qquad
    \subfloat[\centering Mentions]{{\includegraphics[width=0.49\linewidth]{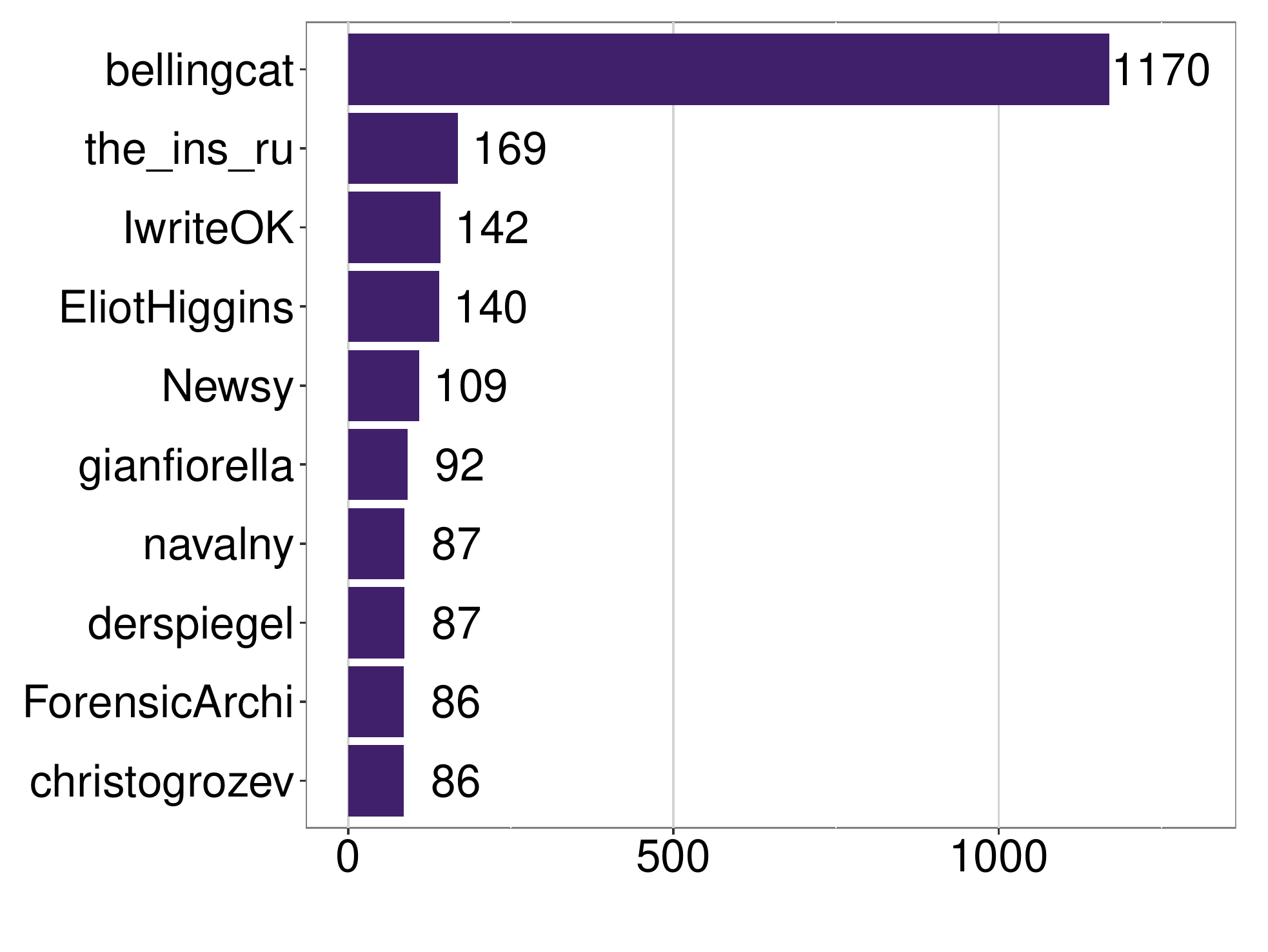}\hfill }}%
    \caption{Top-10 most common (a)~hashtags and (b)~mentions in threads by Bellingcat.}%
    \label{fig:stats_hashtags_mentions}%
\end{figure}

We also study the use of hashtags. Overall, Bellingcat uses 1005 distinct hashtags included in 13\,\% of their threads. The most popular hashtags are shown in Fig.~\ref{fig:stats_hashtags_mentions}(a). Evidently, all popular hashtags refer to an event investigated by Bellingcat such as the downing of Malaysian Airlines flight MH17 (\texttt{\#MH17}), the Russian invasion of Ukraine (\texttt{\#PutinAtWar}), and the poisoning of Sergei Skripal (\texttt{\#TheSalisburyPoisonings}). Hence, Bellingcat strategically uses hashtags to refer to specific events.

We now analyze which accounts are frequently mentioned by Bellingcat on Twitter; see Fig.~\ref{fig:stats_hashtags_mentions}(b). We find that Bellingcat primarily mentions team members such as Elliot Higgins and Giancarlo Fiorella. In some cases, Bellingcat mentions other news outlets (e.g., \texttt{@derspiegel}) and investigative organizations (e.g., \texttt{@ForensicArchi}) as well as external journalists (e.g., \texttt{@ClarkeMikah}). This implies that mentions are used both as a strategy to invite out-group members (potentially with the intention of encouraging reshares) and to promote their own content.  

\noindent
\emph{What topics does Bellingcat discuss on Twitter?} To explore \emph{what} Bellingcat posts about on Twitter, we perform topic analysis (see Methods). Tbl.~\ref{tbl:thread_topics} lists the topics and their corresponding labels. We find that all topics are related to major investigations and projects by Bellingcat. Specifically, seven topics are related to prominent investigations by Bellingcat. Here, 10.90\% of all threads refer to the topic \emph{chemical attacks during the Syrian war}, 10.58\% to the topic \emph{downing of MH17}, and 6.21\% to the topic \emph{Russian invasion of Ukraine}. Given Bellingcat's expertise in uncovering war crimes, this is currently one of the projects pooling large resources \cite{Bellingcat.2022,Pelley.2022}.

In addition to Bellingcat's investigations, we find three topics discussing investigation procedures. The topic \emph{open-source and crowdsourced investigation} captures calls-to-action to Bellingcat's audience on social media for the submission and verification of certain images and videos using collaborative crowdsourcing platforms such as Check (\url{meedan.com/check}). Other topics are related to self-promotion such as \emph{Bellingcat's media announcements} (13.51\%) and \emph{training announcements} (13.30\%). In the latter, Bellingcat offers educational programs on verification techniques, geo-based research tools, image verification, and social media research.

\begin{table*}[ht]
\footnotesize
\centering
\begin{tabular}{r ll r}
\toprule
Topic &  Name & Key terms & Freq. (\%) \\
\midrule
1  & {Chemical attacks during the Syrian war} & attack chemical syria bomb sarin opcw report chlorine syrian russia & 10.90 \\
\addlinespace[0.3em]
2  & {Downing of MH17} & mh russia russian buk ukraine investigation evidence video missile downing & 10.58 \\
\addlinespace[0.3em]
3  & {Geolocation and analysis of Russian’s airstrikes} & video image location russian strike report airstrike showing taken one & 8.55 \\
\addlinespace[0.3em]
4  & {Scripal poisoning} & russian gru russia theinsru fsb investigation one suspect skripal poisoning & 8.51 \\
\addlinespace[0.3em]
5  & {Uncovering global crimes and human right abuses} & navalny team truth poisoning investigation world work europol christogrozev one & 7.79 \\
\addlinespace[0.3em]
6  & {Russian invasion of Ukraine} & newsy investigation russia source tool state attack qanon russian ukraine & 6.21 \\
\addlinespace[0.1em]
7 & {Police violence in Columbia} & police protest rncincle cerosetenta forensicarchi oil wammezz baudoap new press & 5.63\\
\addlinespace[0.3em]
8  & {Open-source and crowdsourced investigation} & mh new report open source eliothiggins investigation kickstarter coming founder & 15.04\\
\addlinespace[0.3em]
9  & {Bellingcat’s media announcements} & podcast episode open ukraine source investigation new also bellingchat patreon & 13.51 \\
\addlinespace[0.3em]
10  & {Bellingcat’s training announcements} & workshop imagery new russian satellite investigation london ymen detail digital & 13.30\\
\bottomrule
\end{tabular}
\caption{Results for topic modeling showing what content is posted by Bellingcat.}
\label{tbl:thread_topics}
\end{table*}

\begin{figure}[H]
    \centering
   {{\includegraphics[width=0.75\linewidth]{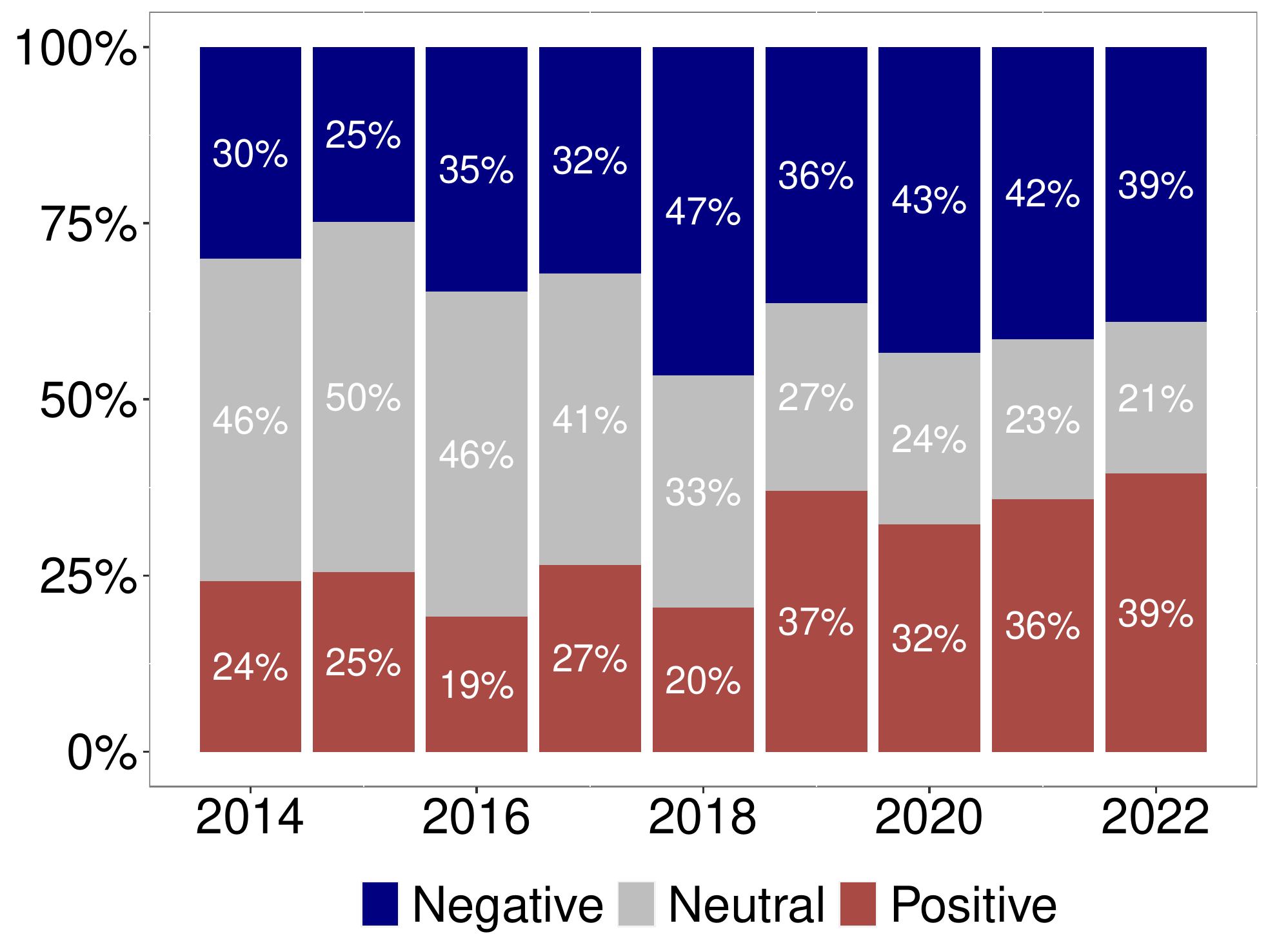}}}%
    \qquad
    \caption{Sentiment embedded in Bellingcat threads over time.}%
    \label{fig:sentiment}%
\end{figure}

\vspace{0.1cm} \noindent
\emph{How positively/negatively does Bellingcat communicate?} To understand \emph{how} positively/negatively Bellingcat communicates, we analyze the valence of Bellingcat's threads using sentiment analysis (see Fig.~\ref{fig:sentiment}). Here, we acknowledge that many topics of Bellingcat feature violence, war crimes, or abuse, rendering it likely that negative sentiment is simply a reflection of the terminology for such events. Interestingly, we observe that, from 2017 onward, Bellingcat has chosen less neutral wording. Hence, Bellingcat started to use more polarized language when it became more popular.  

\subsection{RQ2:What are characteristics of successful social media activities at Bellingcat?}

We study the characteristics of successful social media activities at Bellingcat. To do so, we (1)~compare the engagement of threads featuring media items, hashtags, and mentions to threads without attachments, (2)~compare the engagement of negative, positive, and neutral threads, and (3)~use regression analysis to study how the number of media items, hashtags, mentions, and sentiment is associated with the engagement of Bellingcat's threads. In the following, we define engagement as the sum of likes, quotes, retweets, and replies in a thread, normalized by the number of followers for that month and the thread length. Due to the normalization, we ensure comparability across time and topics. 

\vspace{0.1cm} \noindent
\emph{How to use media items, hashtags, and mentions to promote engagement?} Previous research has shown that media such as images and videos positively influence the success of posts \cite{Hwong.2017, Li.2020b}. Similarly, posts featuring hashtags and mentions reach a broader audience and thus increase engagement \cite{Hwong.2017, Naveed.2011, Suh.2010}. Motivated by this, we examine whether similar effects are found for Bellingcat. For this, we compare threads with vs. without such features and then test for statistically significant differences in the distribution using a Kolmogorov-Smirnov (KS) test \cite{Smirnov.1939}.

\begin{figure*}
\centering
\subfloat[\centering URLs]{{\includegraphics[width=0.18\linewidth]{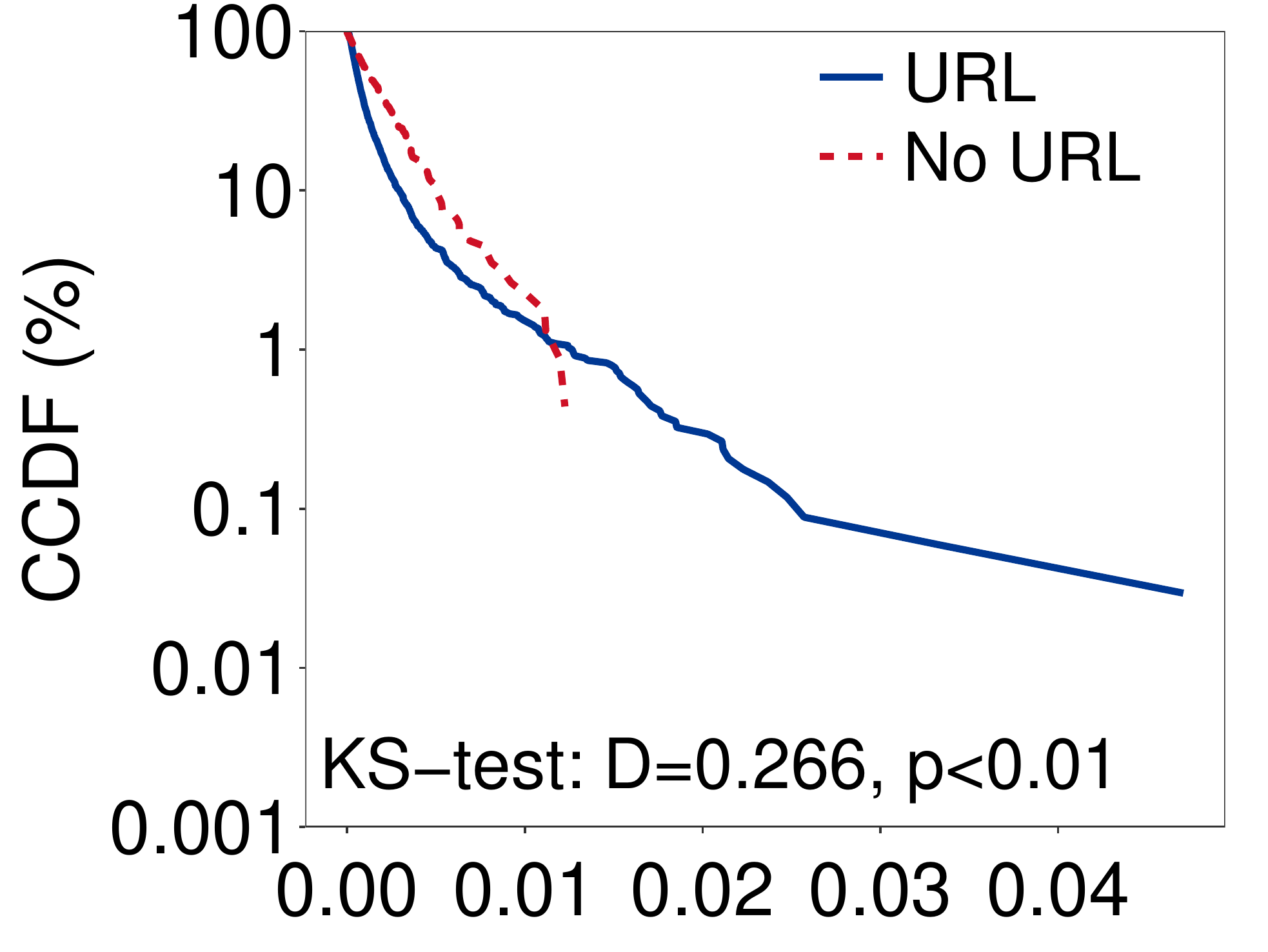}}}
\subfloat[\centering Image ]{{\includegraphics[width=0.18\linewidth]{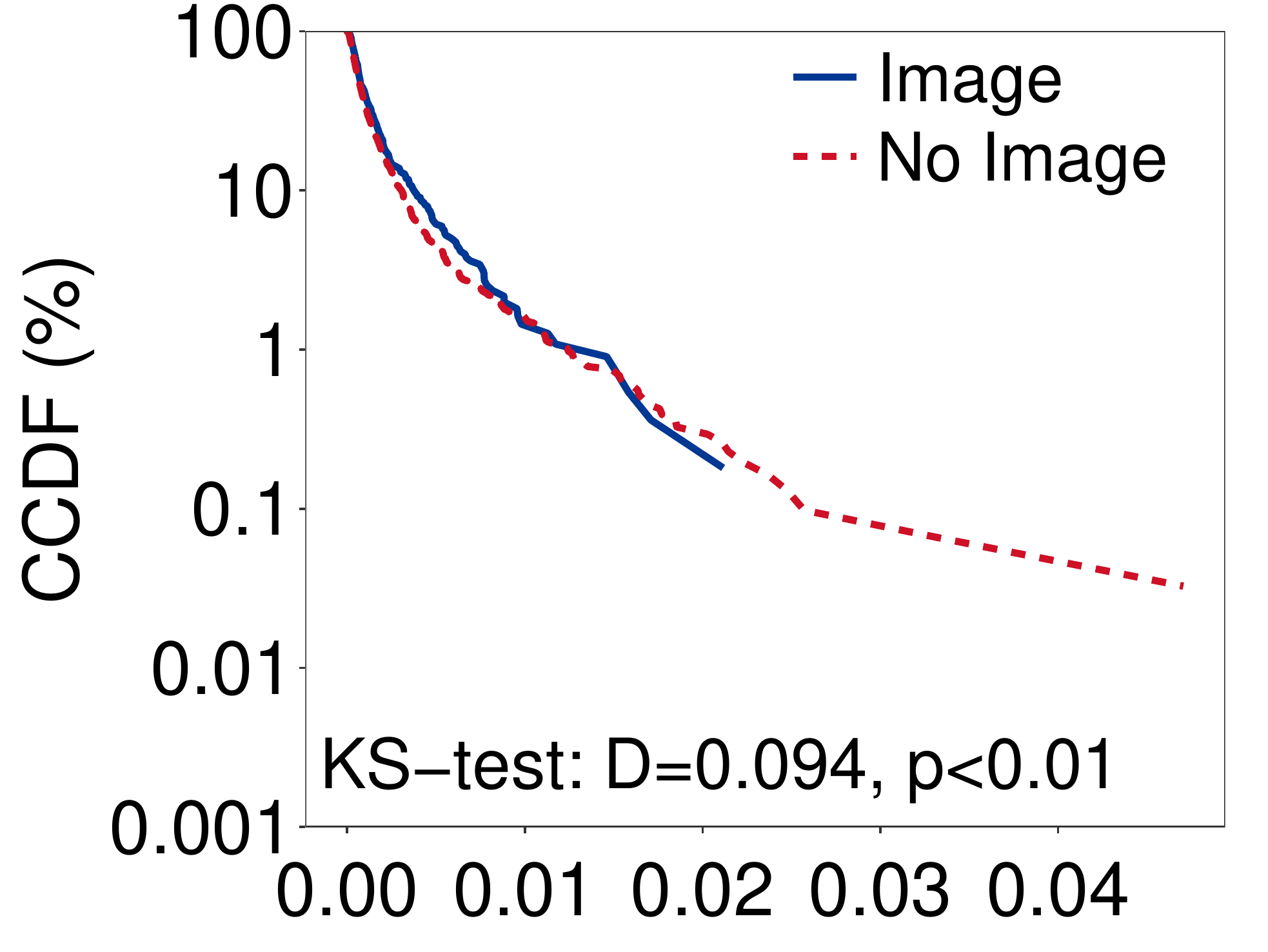}\hfill }}
\subfloat[\centering Video ]{{\includegraphics[width=0.18\linewidth]{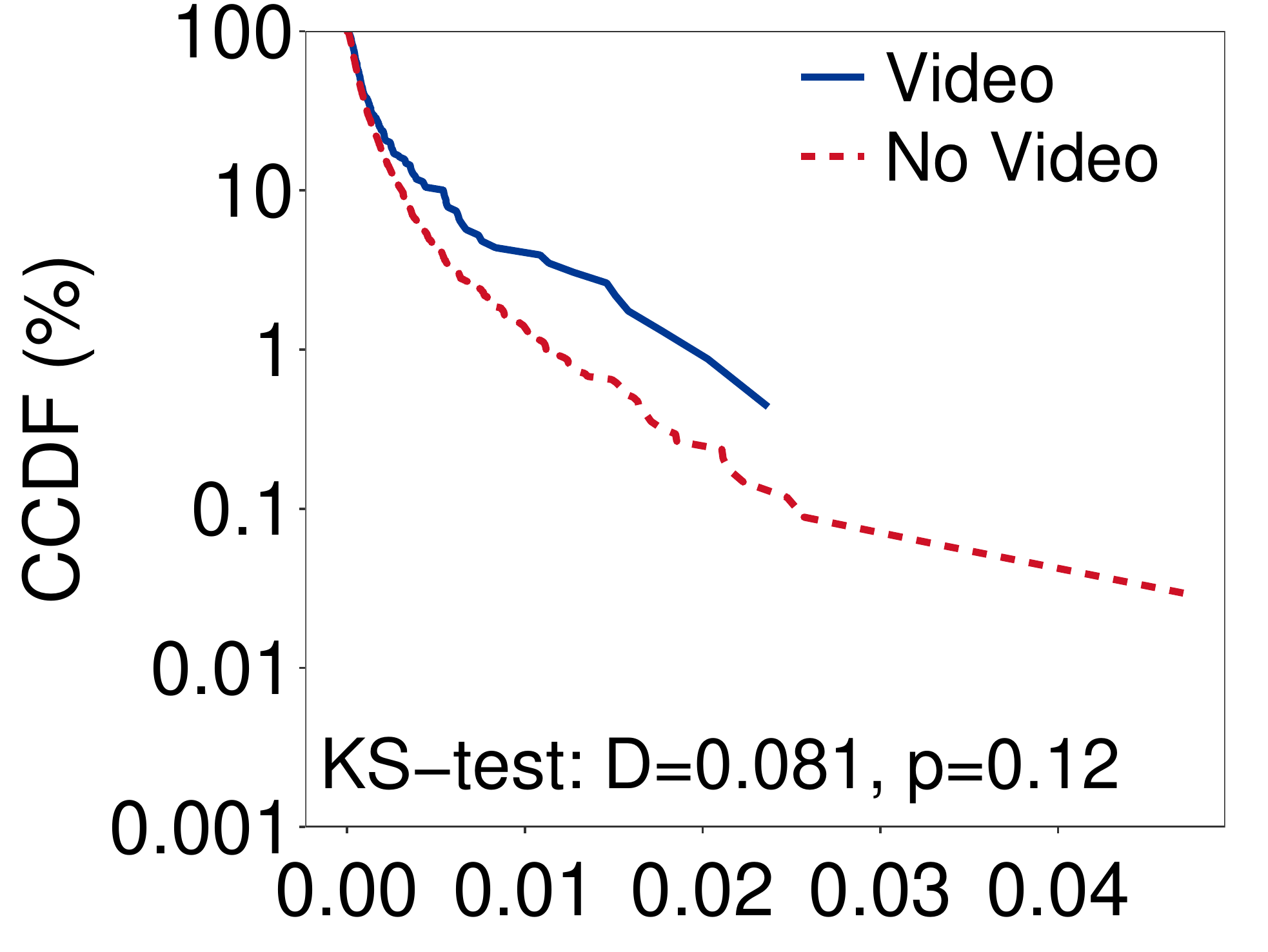}\hfill }}
\subfloat[\centering Hashtags ]{{\includegraphics[width=0.18\linewidth]{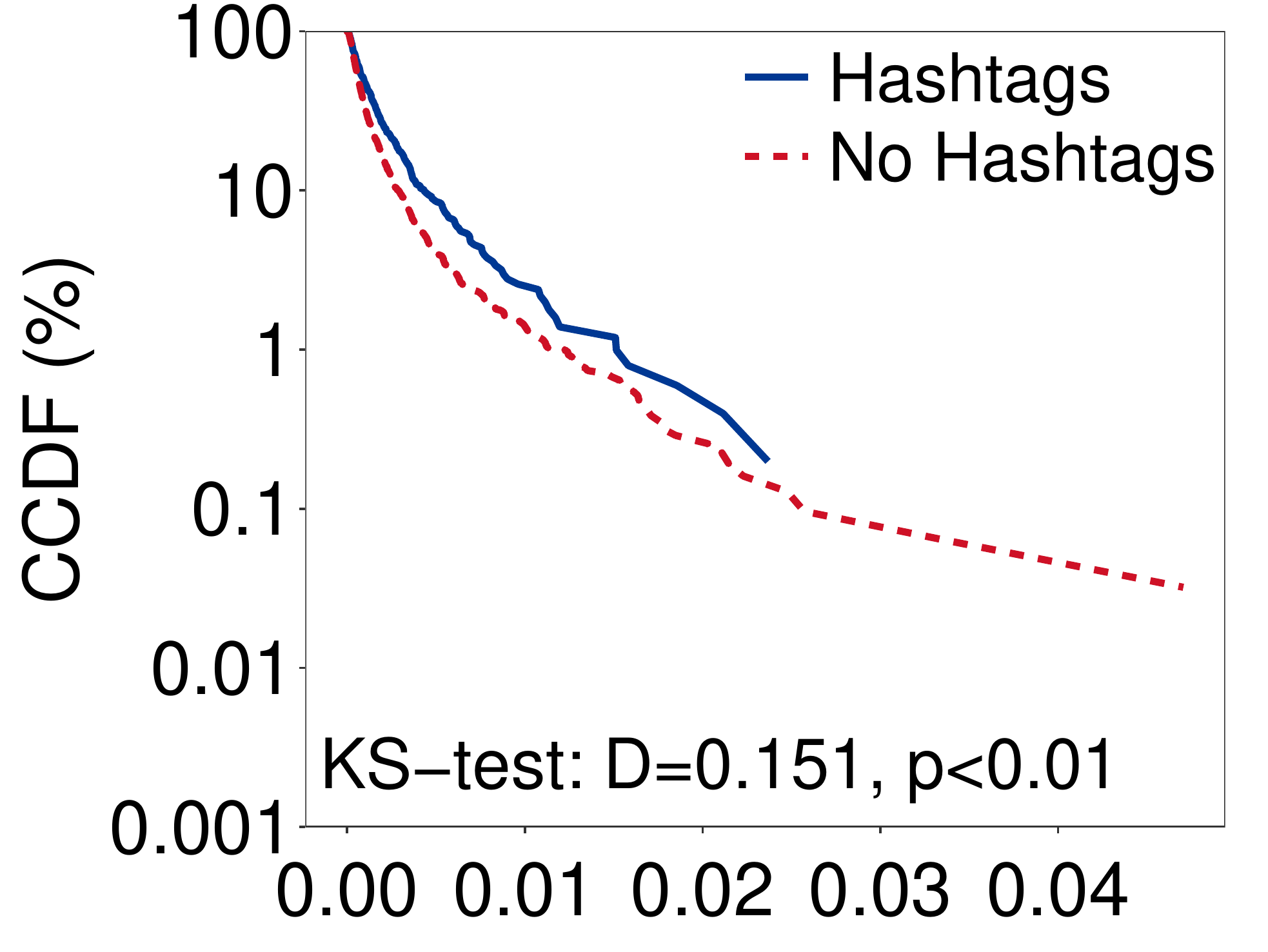}\hfill }}
\subfloat[\centering Mentions ]{{\includegraphics[width=0.18\linewidth]{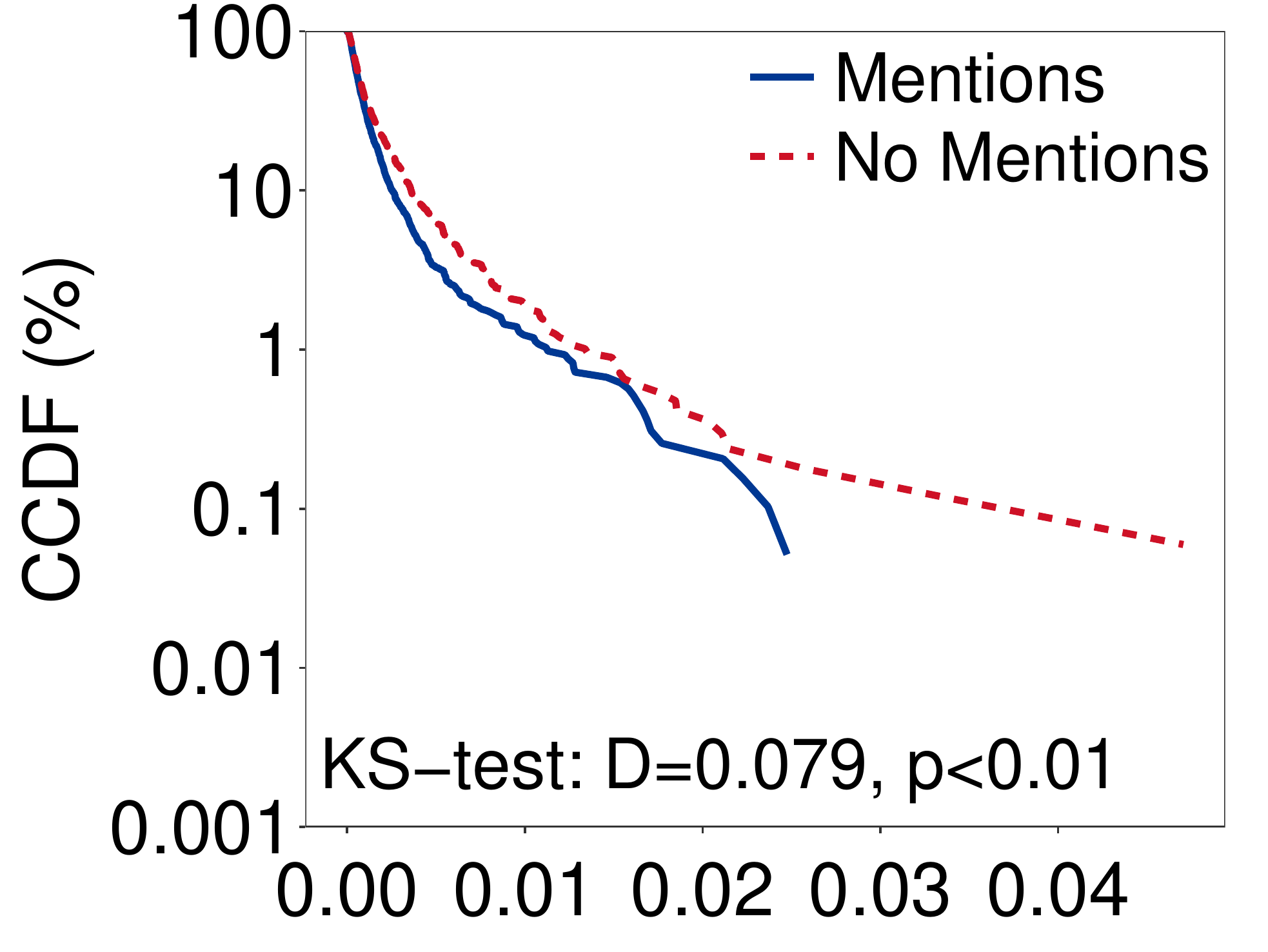}\hfill }}
\caption{Complementary cumulative distribution functions (CCDFs) of engagement with threads that do and do not contain: (a)~URLs, (b)~images, (c)~videos, (d)~hashtags, and (e)~mentions. Plots report results for a Kolmogorov-Smirnov (KS) test to check whether the difference between with vs. without is statistically significant.}
    \label{fig:engagement_media}
\end{figure*}

Fig.~\ref{fig:engagement_media} shows the engagement for threads with different media types (i.e., URLs, images, videos). We find statistically significant differences for threads with URLs, images, hashtags, and mentions (all $p<0.01$). On average, threads with URLs exhibit 53\,\% less engagement while threads with images reach 12\,\% more engagement. Threads that feature hashtags show a 49\,\% higher level of engagement. Further, mentioning other accounts has a 14\,\% lower engagement. For videos, we do not find a statistically significant difference in the distribution of engagement ($p=0.12$), which may be attributed to the low sample size for threads with videos. In sum, Bellingcat can promote engagement through the use of images and hashtags but not through mentions or links to other websites.

\begin{figure}
    \centering
    \includegraphics[width=0.65\linewidth]{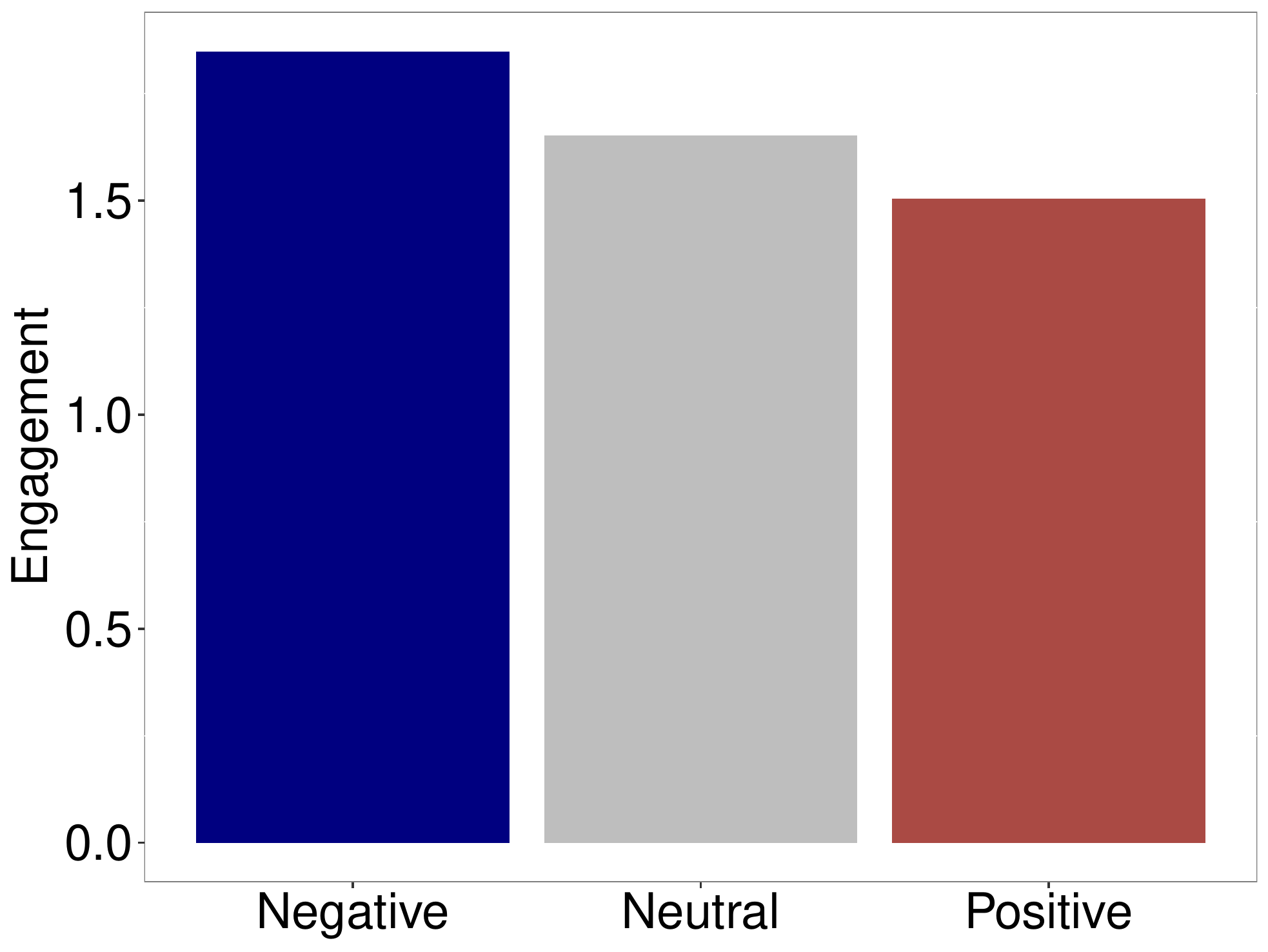}
    \caption{Engagement (defined as the sum of likes, quotes, retweets, and replies in a thread, normalized by the number of followers for that month and the thread length) for threads embedding negative, neutral, or positive language.}
    \label{fig:engagement_sentiment}
\end{figure}

\noindent
\emph{Which sentiment elicits user engagement?} Previous research showed that negative language can increase engagement with news and other content online \cite{Hansen.2011,Jenders.2013,Robertson.2022}. In line with a negativity bias, it is interesting to test whether negative threads also reach a larger audience. Fig.~\ref{fig:engagement_sentiment} compares the engagement for threads with negative, neutral, and positive sentiment. We find that threads with negative sentiment are characterized by higher levels of engagement compared to threads with neutral and positive sentiment. Specifically, aggregated engagement is 12\,\% (22\,\%) higher for threads featuring negative language compared to threads with neutral (positive) language. As such, threads by Bellingcat are more successful when they are framed with a negative wording.

\vspace{0.1cm} \noindent
\emph{How are thread characteristics associated with engagement?} Above, we showed that engagement with threads depends on various characteristics. Hence, we now provide a statistical analysis to study how thread characteristics are associated with different levels of engagement. To do so, we estimate the following regression model
\begin{multline}\label{eqn:regression}
    log(Engagement_i) = \beta_0 + \beta_1URL_i + \beta_2Image_i + \beta_3Video_i \\ 
    + \beta_4Hashtags_i + \beta_5Mentions_i + \beta_6Sentiment_i ,
\end{multline}
where we regress the number of media items (i.e., URLs, images, videos), hashtags, and mentions as well as the sentiment of a thread on engagement. As engagement is heavily skewed, we log-transformed our dependent variable.\footnote{We repeated our analysis without log-transformation of engagement, which resulted in qualitatively similar results. For details, see our GitHub.} For estimation, we use ordinary least squares regression (OLS) and tested whether the coefficients are significantly different from zero using two-sided $t$-tests.

The regression results are in Tbl.~\ref{tbl:regression}. We find a positive and statistically significant coefficient for images ($P<0.01$) and videos ($P<0.01$). Hence, threads that contain images or videos spark higher levels of engagement. In contrast, we find a negative and statistically significant relationship of URLs with engagement ($P<0.01$). Evidently, visual information is more successful compared to content from other websites. Furthermore, we find a positive coefficient for hashtags and a negative coefficient for mentions and sentiment. While these coefficients are not statistically significant, the results are in line with our descriptive analysis. Therefore, they indicate that threads are more successful if they contain more hashtags, fewer mentions, and are written with a negative sentiment.

\begin{table}[!ht]
\centering
\footnotesize
\begin{tabular}{lS[table-format = 2.3]S[table-format = 1.4]S[table-format = 2.3]S[table-format = 1.3]r}
\toprule
& {Coef.} & {SE} & {$t$-value} & {$P$-value} & {95\,\%~$\text{CI}$} \\ 
\midrule
Intercept         &      -7.2379  &        0.031     &  -230.866  &         0.000        &       $[-7.299, -7.176]$     \\
URL          &      -0.0472  &        0.016     &    -2.918  &         0.004        &       $[-0.079, -0.015]$     \\
Image  &       0.0673  &        0.024     &     2.835  &         0.005        &        $[0.021, 0.114]$     \\
Video &       0.1928  &        0.073     &     2.647  &         0.008        &        $[0.050, 0.336]$     \\
Hashtags      &       0.0456  &        0.095     &     0.481  &         0.630        &       $[-0.140, 0.231]$     \\
Mentions     &      -0.0107  &        0.009     &    -1.174  &         0.240        &       $-0.028, 0.007]$     \\
Sentiment      &      -0.0535  &        0.040     &    -1.337  &         0.181        &       $[-0.132, 0.025]$     \\
\bottomrule
\multicolumn{6}{l}{\tiny{SE = standard error, CI = confidence intervals}} \\
\multicolumn{6}{l}{\tiny{Standard errors are heteroscedasticity robust}}
\end{tabular}
\caption{Regression results to study how thread characteristics drive engagement.}
\label{tbl:regression}
\end{table}

\subsection{RQ3: How did the social media activities on Bellingcat’s Twitter channel change in response to the Russian Invasion of Ukraine?}
 
Social media has played a salient role in the discourse of the Russian invasion of Ukraine \cite{Geissler.2022}. For example, data from Reddit shows that conversations resemble those of Russian disinformation websites \cite{Hanley.2022}. Furthermore, the NATO Strategic Communication Center considers social media as highly influential and states that Russia exploits social media websites to spread propaganda \cite{Baack.2015}. One assumption is that this has shifted both the overall audience (e.g., by attracting new followers), investigations (e.g., Ukraine now pools extensive resources), and social media strategies (e.g., to provide timely fact-checks and debunk Russian disinformation) of Bellingcat. Motivated by this, we now analyze how social media activities have changed in response to the war. To offer statistical comparisons, we focus on the time frame $\pm$100 days pre/post the Russian invasion of Ukraine on February 24th, 2022.

\begin{figure}[H]
    \centering
    \includegraphics[width=0.75\linewidth]{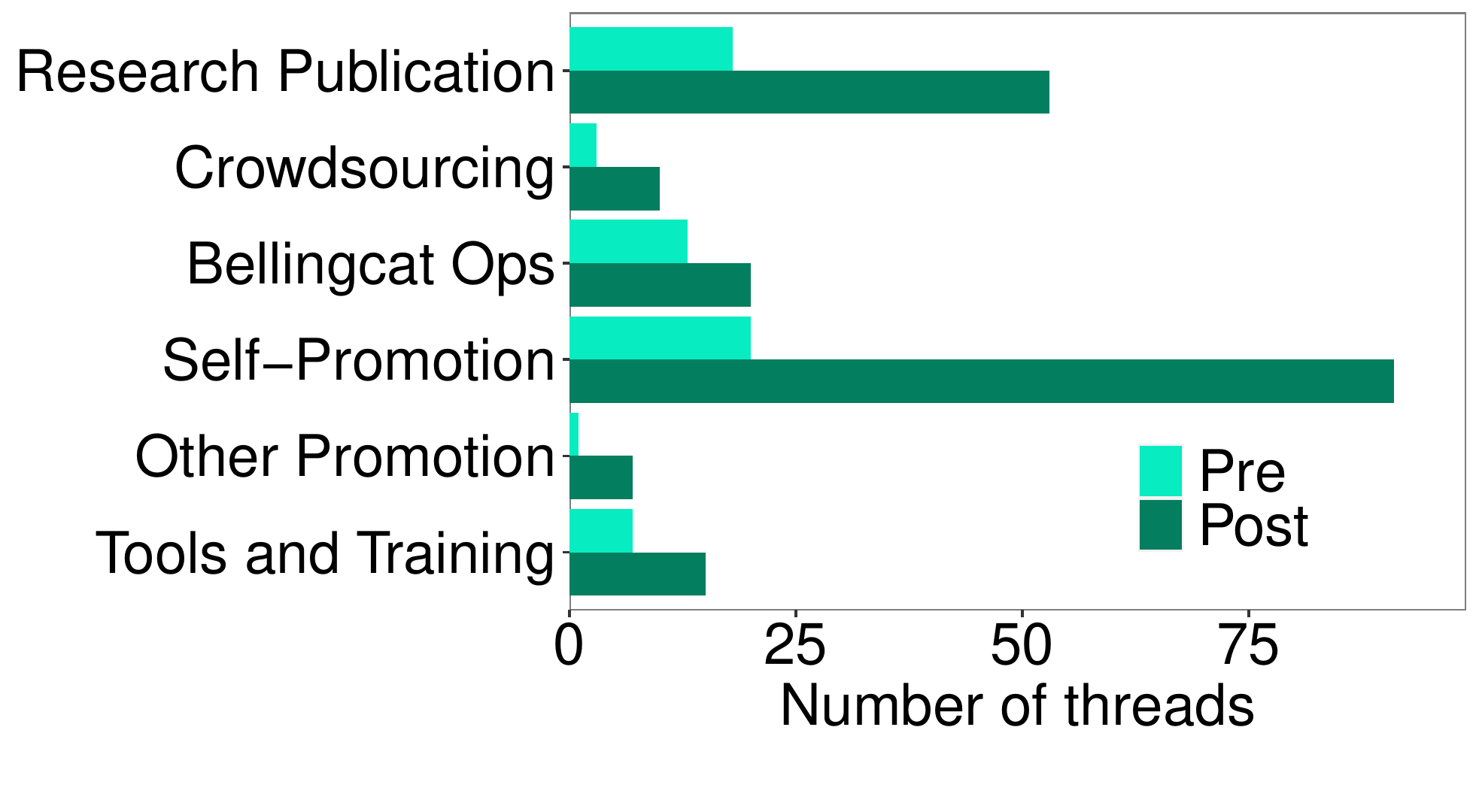}
    \caption{Change in the number of threads by purpose pre/post the Russian invasion of Ukraine ($\pm100$ days).}
    \label{fig:change_volumen_war}
\end{figure}
  
\vspace{0.1cm} \noindent
\emph{How does the narrative of social media content change after the Russian invasion of Ukraine?} We inferred the purpose of all threads posted by Bellingcat in the 100 days before and after the Russian invasion of Ukraine using our qualitative classification by purpose (see Methods). The results are displayed in Fig.~\ref{fig:change_volumen_war}. We observe a significant increase in the number of threads posted by Bellingcat for all purpose categories. This increase is particularly pronounced for the purpose categories \textsc{Other-Promotion}, \textsc{Self-Promotion}, and \textsc{Crowdsourcing}. Here, the number of threads related to \textsc{Self-Promotion} increased by 355\,\% which corresponds to 46\,\% of all threads post-invasion. For the purpose category \textsc{Other-Promotion}, we find an even larger increase  of 600\,\% post-invasion. Overall, this highlights that Bellingcat puts additional efforts into promoting the organization but also shares other content to interact with their community and spawn crowdsourcing. 

\begin{figure}[H]
    \centering
    \includegraphics[width=0.5\linewidth]{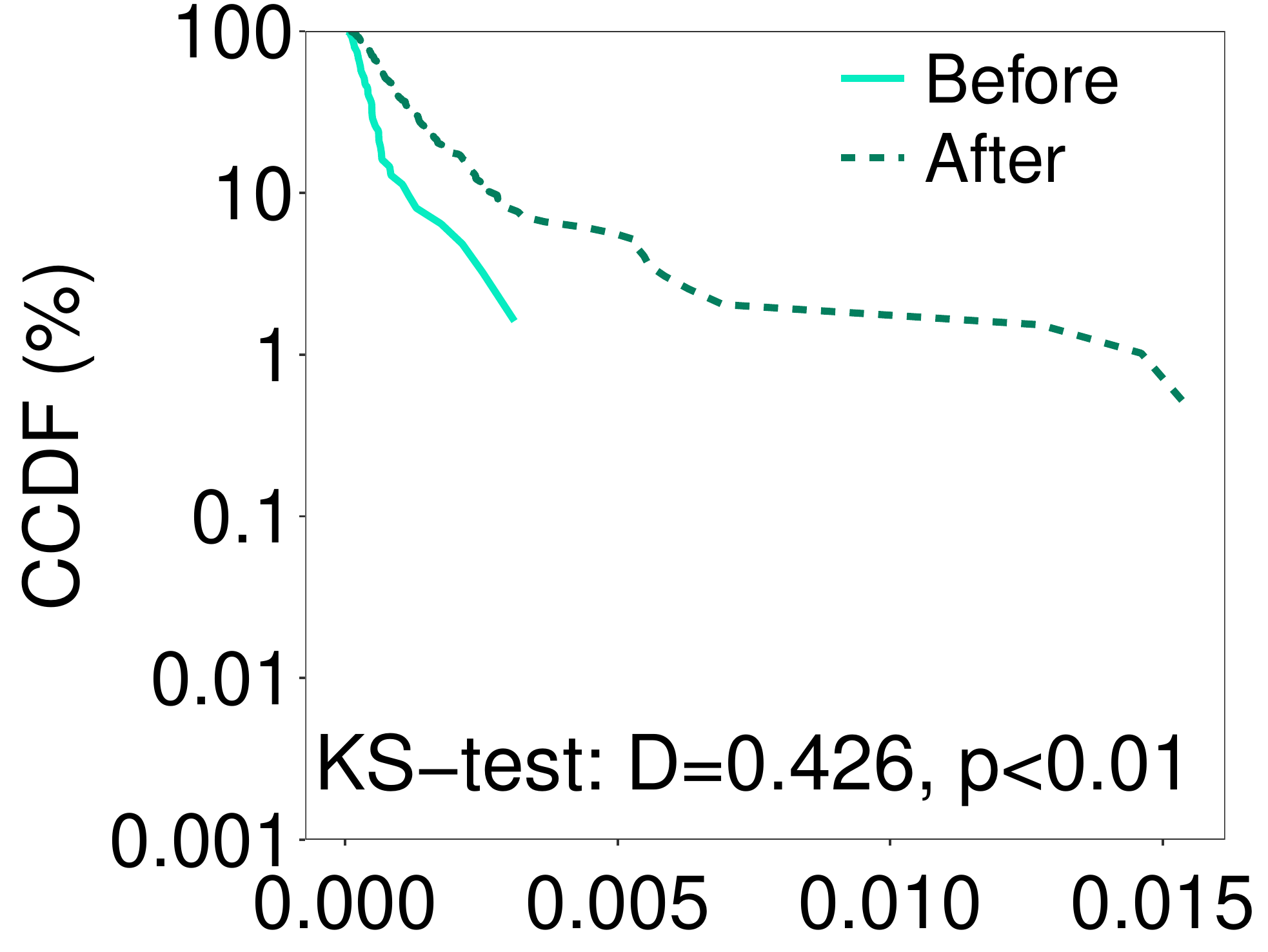}
    \caption{Engagement for threads pre/post the Russian invasion of Ukraine ($\pm100$ days). Engagement is defined as the sum of likes, quotes, retweets, and replies in a thread, normalized by the number of followers for that month and the thread length. Kolmogorov-Smirnov (KS) test shows that the difference between before vs. after is statistically significant.}
    \label{fig:engagement_war}
\end{figure}

 \vspace{0.1cm} \noindent
\emph{How does engagement behavior change in response to the Russian invasion of Ukraine?} We now study the engagement of threads posted by Bellingcat pre- vs. post-invasion. We expect that the prominence of this topic for society also leads to higher levels of engagement, especially since Bellingcat puts increased efforts into investigations related to the Russian invasion of Ukraine. We find that threads after the invasion have a significantly different engagement distribution ($p<0.01$); see Fig.~\ref{fig:engagement_war}. Also, they exhibit a statistically significant increase in engagement ($p<0.01$) as demonstrated by a Mann-Whitney U-test. This shows that threads posted by Bellingcat after the Russian invasion of Ukraine receive considerably more likes, replies, and retweets. This is in line with information value theory, which states that topics perceived as ``high value'' (e.g., content with negative consequences and potentially unexpected findings) increase engagement on social media \cite{Rudat.2015}. 

\vspace{0.1cm} \noindent
\emph{How does the Russian invasion of Ukraine affect replies to Bellingcat's threads?} We saw that engagement with threads by Bellingcat post-invasion increased significantly. To gain an even better understanding of how users interact with Bellingcat, we now study how replies to threads posted by Bellingcat have changed post-invasion. For this, we compare the number of replies as well as URLs, visual media (i.e., images and videos), and hashtags shared along with replies. Furthermore, we analyze how the valence of the replies changes pre- and post-invasion. 

The results are in Tbl.~\ref{tbl:replies} pointing to a change in user replies after the invasion. In particular, for post-invasion, a thread receives three times as many replies as threads posted by Bellingcat pre-invasion. Additionally, there is an increase in the number of URLs, visual media, and hashtags shared with replies. This is particularly pronounced for hashtags, which almost double.

\begin{table}[H]
\centering
\footnotesize
\begin{tabular}{lllll}
\toprule
Time frame &  Replies & URLs & Media items & Hashtags \\
\midrule
Pre war & 22.64 & 0.135 & 0.052 & 0.041 \\
Post war & 60.73 & 0.180 & 0.080 & 0.080 \\
\bottomrule
\end{tabular}
\caption{Summary statistics (as mean per thread) comparing threads pre/post the Russian invasion of Ukraine ($\pm100$ days).}
\label{tbl:replies}
\end{table}

\vspace{0.1cm} \noindent
\emph{How does content creation change after the Russian invasion of Ukraine?} To analyze changes in content creation, we analyze the top~20 hashtags in the replies which should provide a proxy for the topics discussed. We find that the replies featured a broad variety of hashtags pre-invasion (e.g., \texttt{\#visituganda} and \texttt{\#swissarms}) but the top-20 hashtags post-invasion refer exclusively to the Russian invasion of Ukraine (e.g., \texttt{\#Ukraine} and \texttt{\#StandWithUkraine}). As such, the replies reflect Bellingcat's shift to investigations around the Russian invasion of Ukraine.

\begin{figure}[H]
\centering
\includegraphics[width=0.65\linewidth]{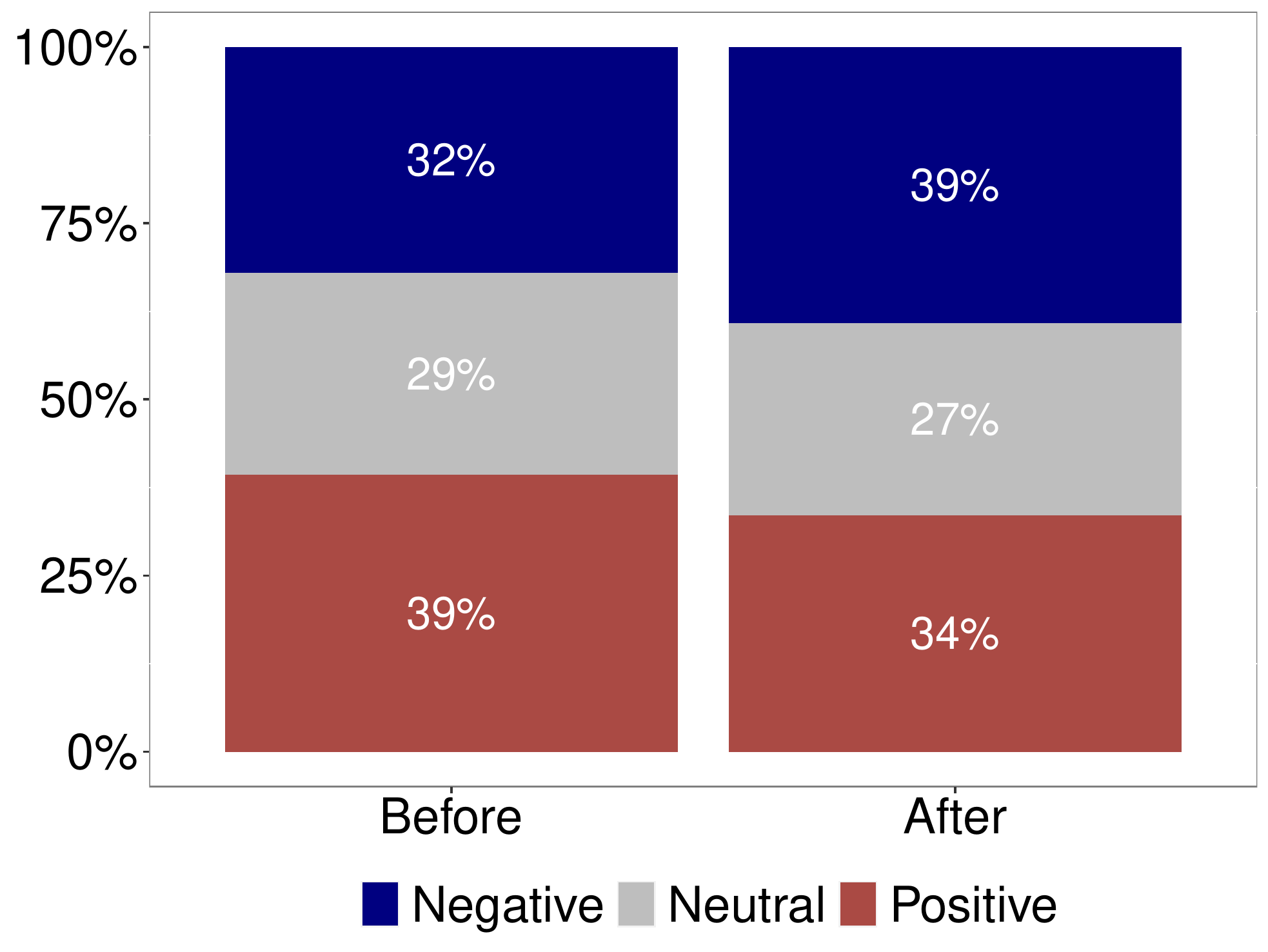}
\caption{Impact of the Russian invasion on sentiment in user replies. The comparison is based on $\pm100$ days pre/post the Russian invasion of Ukraine.}
\label{fig:change_sentiment_war}
\end{figure}

Previous research has shown that users react with different sentiment to crisis events \cite{Jakubik.2022}. Hence, we compare the sentiment of replies to Bellingcat's threads pre- vs. post-invasion (see Fig.~\ref{fig:change_sentiment_war}). Our results show a larger share of negative replies post-invasion as compared to pre-invasion. In particular, the share of replies featuring negative sentiment increased by 7.13\,\%, while the share of positive replies decreased by 5.76\,\%. Overall, this implies that replies to threads by Bellingcat became more negative in light of the Russian invasion of Ukraine.

\vspace{0.1cm} \noindent
\emph{What is the impact of bots on Bellingcat's Twitter account?} Due to Bellingcat's efforts to investigate Russian war crimes after the Russian invasion of Ukraine, Bellingcat's account might be prone to Russian propaganda in the form of bots. Notwithstanding, Russia is known for a long tradition of initiating disinformation campaigns through bot-led networks \cite{Bail.2020, Balasubramanian.2022}.
Hence, we now analyze the impact of bots on Bellingcat's Twitter account.  

We used Botometer to identify bots replying to threads by Bellingcat (see Methods). We find that the number of bots increased substantially post-invasion. While we only found 93 bots pre-invasion, we found 1,279 bots replying to Bellingcat post-invasion, which is a $\sim$13-fold increase. Overall, these bots posted 421 replies pre-invasion compared to 2,655 replies post-invasion. We also compared the use of URLs, visual media, and hashtags by bots (not shown for reasons of space). Here, we found that bots use relatively fewer attachments which is contrary to our findings above. A reason might be a higher degree of bot automation, which is facilitated by fewer attachments to replies. 

\section{Discussion}

\textbf{Relevance:} OSJ is a new phenomenon in the media ecosystem aimed at producing investigations from open-source information and, thereby, holding governments and other organizations accountable for misconduct \cite{Muller.2021}. Social media is crucial for OSJ, such as Bellingcat, since, unlike  mainstream media outlets, social media is used for crowdsourcing, and, further, it presents the main channel for distributing reports \cite{Cooper.2021, Russell.2019}. To the best of our knowledge, our work is the first to characterize the social media activities of OSJ.
 
\textbf{Main findings:} Our results identify distinctive characteristics of social media activities at Bellingcat that are different from traditional journalistic organizations. For instance, OSJ does not solely rely on social media for sharing publications of their investigations or self-promotion, but they actively engage with their audience in crowdsourcing information. Here, we found that $\sim$5\,\% of the social media posts fall under this. Previous research has studied crowdsourcing in various settings, such as for natural disaster relief \cite{Gao.2011}, prediction markets \cite{Freeman.2017}, or funding \cite{Lu.2014}, whereas we add by crowdsourcing for investigative intelligence. In addition to crowdsourcing, Bellingcat frequently launches tweets to encourage participation. For example, tools and training for open-source intelligence are frequently promoted ($\sim$8.5\,\%). This is not surprising since the recurrent theme at Bellingcat is to rely on ``open-source'' and bring their tools to other actors and especially to other parts of the world. Further, different from traditional media outlets, reports from others and not from Bellingcat are actively shared and discussed ($\sim$3\,\%). 

Our work identifies levers that elicit engagement for OSJ on social media. We found user engagement is particularly large for posts with a negative sentiment. This finding is in line with theory on the negativity bias, according to which negativity draws attention and thus drives user behavior \cite{Berger.2012,Berger.2011,Soroka.2019}. Similar observations have been made by prior research in the case of news where negative language increases consumption \cite{Robertson.2022} and sharing \cite{Hansen.2011, Jenders.2013,Naveed.2011,Stieglitz.2013}. Moreover, we found that the use of images also promotes user engagement. This finding supports other literature, where visual content on Twitter is generally very engaging \cite{Hwong.2017, Li.2020b}.  

\textbf{Implications:} Our work highlights several implications for OSJ. First, we provide recommendations for effective social media activities, such as on how to elicit engagement. One example is impression management which leverages negativity bias to draw attention. Here, prior research has found that, at the news platform ``Upworthy'', headlines with a negative sentiment received more clicks \cite{Robertson.2022}. Another example is the use of images, which is consistent with prior theory that visual cues drive human attention \cite{Hwong.2017, Li.2020b}.

Second, our results highlight the importance of collective actions that involve the crowd for successful social media activities. Here, our results document that the follower base can be leveraged not only for effective information dissemination (as in the case of traditional media outlets) but also for the purpose of information gathering through crowdsourcing. This may be an interesting avenue for traditional media outlets to expand their research efforts.

Third, we showed that major world events -- as is the case for the Russian invasion of Ukraine -- can have a significant impact on social media activities. For Bellingcat, we see a substantial increase in both the number of followers and their engagement. Moreover, a particularly alarming observation is that we also found a significant increase in bot activities. OSJ organizations must be aware of this and should establish mechanisms to mitigate potential negative effects. Importantly, bots may not only interfere with the regular social media activities at OSJ but, if they post comments with disinformation or propaganda, may undermine the efforts of OSJ to present evidence-based facts. Eventually, this could raise the risk of promoting false beliefs among followers, contrary to the original intentions of OSJ. Hence, it is important for OSJ to define scalable ways to effectively moderate the discussion around sensitive topics, even in the presence of bots and disinformation campaigns.

\textbf{Ethical Considerations:} Our results have direct ethical implications. For example, communication strategies that leverage negativity bias may spur negative reactions by other users \cite{Cheng.2017, Shmargad.2022} and trigger conflicts \cite{Kumar.2018} which can deteriorate communication on social media and the offline world. Consequently, OSJ must consider the online and offline implications of their communication to avoid negative social outcomes. Responsible communication is particularly important in today's online ecosystem where high levels of misinformation and fake news already erode trust in mainstream institutions and news media \cite{Bar.2023,Ognyanova.2020}.

\textbf{Limitations and future research:} As with others, our work is subject to limitations that open avenues for future research. First, our analysis focused on Bellingcat’s use of Twitter due to the large follower base ($\sim$709,900 followers). We acknowledge that Bellingcat is also active on other social media platforms, but the follower base is significantly smaller in size (e.g., their Telegram channel has $\sim$31,000 members and their Facebook page has $\sim$81,000 followers) and other platforms receive less coverage as only a subset of the Twitter content is posted there. Second, we focused on the official Bellingcat account but are aware that individual journalists from the collective operate their own accounts. Hence, future research may also study the social media activities of individual journalists. Nevertheless, the official Bellingcat account presents the main source for publications, and we expect our results to be transferable. Third, our results are of associative nature and do not lend to causal interpretations. For example, external events generally lead to more activity on Twitter such that Bellingcat's growth in followers and interactions may not be a direct cause of their investigations. However, this is a common limitation in settings where randomized controlled trials are challenging or impossible \cite{Maarouf.2022, Prollochs.2022}.

\section{Conclusion}

Open-source journalism emerged as a new phenomenon that leverages crowdsourcing using public data to produce investigations. In this paper, we characterize the social media activities at Bellingcat on Twitter. We found that Bellingcat's growth on Twitter is largely influenced by external events such as the Russian invasion of Ukraine. Moreover, we found that a large share of threads (15\,\%) are related to open-source and crowdsourcing investigations. We also analyzed determinants that drive user engagement. Here, threads with images and negative sentiment are characterized by higher levels of engagement. Finally, we provided evidence that the Russian invasion of Ukraine had a significant impact on social media activities: we observed a large increase in followers, replies, and in bot activities. Overall, our findings provide recommendations for how OSJ such as Bellingcat can successfully operate in the media ecosystem.

\section{Ethics Statement}

Data collection was conducted following standards for ethical research \cite{Rivers.2014}. All data used in this study was public and open-source. We respect the privacy and agency of all people potentially impacted by this work and take specific steps to protect their privacy. We do not identify users that have chosen to remain anonymous. We further report data only at the aggregate level. Since this research did not involve interventions, no ethics approval was required by our Institutional Review Board.

\bibliographystyle{ACM-Reference-Format}
\bibliography{bibliography}

\end{document}